\documentclass[12pt, table]{article}
\usepackage{geometry}  % simplest to just add NJP size (can be improved...)
\usepackage{times}
\usepackage[intlimits]{amsmath}
\usepackage{amssymb}
\usepackage{graphicx}
\usepackage{tikz}
\usepackage{booktabs}
\usepackage[colorlinks]{hyperref}
\usepackage{xcolor}
\hypersetup{colorlinks,
    linkcolor={red!20!black},
    citecolor={blue!20!black},
    urlcolor={blue!20!black}
}

\usepackage[utf8]{inputenc}
\usepackage[greek,english]{babel}
\usepackage{mathtools}
\usepackage{multirow}

\graphicspath{{figures/}}

\definecolor{PairedA}{RGB}{166, 206, 227}
\definecolor{PairedB}{RGB}{ 31, 120, 180}
\definecolor{PairedC}{RGB}{178, 223, 138} 
\definecolor{PairedD}{RGB}{ 41, 128,  35}   % made darker by 20%
\definecolor{PairedE}{RGB}{251, 154, 153}
\definecolor{PairedF}{RGB}{182,  21,  22}   % made darker by 20%
\definecolor{PairedG}{RGB}{253, 191, 111}
\definecolor{PairedH}{RGB}{255, 127,   0}
\definecolor{PairedI}{RGB}{202, 178, 214}
\definecolor{PairedJ}{RGB}{106,  61, 154}
\definecolor{PairedK}{RGB}{255, 255, 153}
\definecolor{PairedL}{RGB}{177,  89,  40}

\usepackage[most]{tcolorbox}
\colorlet{ComplexA}{PairedC!25}
\colorlet{ComplexB}{PairedA!25}
\colorlet{ComplexC}{PairedE!25}

\usepackage[colorinlistoftodos,prependcaption,textsize=small]{todonotes}
\colorlet{colorMG}{blue}
\colorlet{colorKS}{red}
\colorlet{colorMC}{green}

\usepackage{mathCommands}

\begin{document}
\title{Upper bounds on absorption and scattering\footnote{This work resulted from Brat\v{r}ice Optimal Optics Meeting (BOOM) that was held on September 3-6, 2019, in Brat\v{r}ice, Czech Republic. For more information see \url{https://elmag.fel.cvut.cz/ESAworkshops}}}
\author{Mats Gustafsson$^1$, Kurt Schab$^2$, Lukas Jelinek$^3$, and Miloslav Capek$^3$}
\date{%
    \small{$^1$ Dept. of Electrical and Information Technology,
Lund University, Lund, Sweden\\%
    $^2$ Dept. of Electrical Engineering, Santa Clara University, Santa Clara, USA\\
    $^3$ Dept. of Electromagnetic Field, Czech Technical University in Prague, Prague, Czech Republic\\[2ex]}%
    \normalsize{\today}
}
%\author{Brat\v{r}ice Optimal Optics Meeting (BOOM)\footnote{Brat\v{r}ice Optimal Optics Meeting (BOOM) was held on September 3-6, 2019, in Brat\v{r}ice, Czech Republic. For more information see \url{https://elmag.fel.cvut.cz/ESAworkshops}.}}
\maketitle
\tableofcontents

\begin{abstract}
A general framework for determining fundamental bounds in nanophotonics is introduced in this paper. The theory is based on convex optimization of dual problems constructed from operators generated by electromagnetic integral equations. The optimized variable is a contrast current defined within a prescribed region of a given material constitutive relations. Two power conservation constraints analogous to the optical theorem are utilized to tighten the bounds and to prescribe either losses or material properties. Thanks to the utilization of matrix rank-1 updates, modal decompositions, and model order reduction techniques, the optimization procedure is computationally efficient even for complicated scenarios. No dual gaps are observed. The method is well-suited to accommodate material anisotropy and inhomogeneity. To demonstrate the validity of the method, bounds on scattering, absorption, and extinction cross sections are derived first and evaluated for several canonical regions. The tightness of the bounds is verified by comparison to optimized spherical nanoparticles and shells. The next metric investigated is bi-directional scattering studied closely on a particular example of an electrically thin slab. Finally, the bounds are established for Purcell's factor and local field enhancement where a dimer is used as a practical example.
\end{abstract}

\section{Introduction}
As the field of nanophotonics becomes more mature, interest is shifting away from the analysis of simple systems (uniform waveguides, spheres, rods, etc.) and toward the synthesis of structures with engineered electromagnetic behavior, \eg{} maximal absorption~\cite{molesky2019fundamental}, directional emission~\cite{hancu2013multipolar,coenen2014directional}, directed scattering~\cite{liu2014ultra,callewaert2018inverse}, fluorescence diplexing~\cite{vercruysse2014directional}, waveguide power division~\cite{Miller+etal2016}, field confinement~\cite{lin2016cavity}, and waveguide diplexing (wavelength splitters)~\cite{piggott2015inverse}.  In order to generate novel geometries optimized for these particular design objectives, computational inverse design is often employed, see, \eg{}~\cite{molesky2018inverse,BendsoeSigmund_TopologyOptimization} and the references therein.  While such methods excel in the exploration of extremely broad design spaces and the discovery of non-intuitive solutions, there nevertheless exists a strong need for analytic results which inform, direct, and truncate their computationally intensive calculations~\cite{molesky2018inverse}.

Physical bounds on performance objectives constitute a particular class of analytic results that aid inverse design in this way.  For example, rather than searching for a design which achieves a particular objective value, one may instead employ inverse design tools to search for an \emph{optimal design} with superior performance over all other possible structures satisfying some fixed constraints.  In searching for this optimal design, the determination of physical bounds is critical in comparing realized performance to the theoretical optimum.  Additionally, secondary analytic results and features of optimal designs may often be obtained through the process of deriving physical bounds, as in the study of optimal antennas operating in the microwave regime. These secondary results add understanding to the behavior of optimal designs and may provide valuable input to specialized inverse design algorithms, \eg{} starting points or feature definitions.

Several approaches have previously been applied to calculate physical bounds on the performance of nanophotonic structures, though each of these methods has unique limitations.  One approach is to analyze the broadband characteristics of scatterers using passivity-based sum rules~\cite{Sohl+etal2007a}.  Sum rules of this kind lead to important conclusions regarding many practical applications (\eg{} cloaking~\cite{Monticone+Alu2013}), though they fail to offer insight into performance bounds at single frequencies. Attempts to combine sum rules with other methods of calculating bounds rely on assuming simple frequency domain responses, \eg{} Lorentzian line shapes~\cite{Shim+etal2019}. Alternatively, analysis may be restricted to canonical structures, such as layered spheres (``core-shell structures''), where the low number of degrees of freedom allows for tight brute force optimization~\cite{Sheverdin+Valagiannopoulos2019}. Dipolar interaction for electrically small structures is analyzed in~\cite{Radi+Tretyakov2013}, spherical mode expansions in~\cite{Liberal+etal2014}, and effects of lossy background media in~\cite{Ivanenko+etal2019}.

Based primarily on the optical theorem, shape independent bounds exist~\cite{Miller+etal2016} assuming uniform fields and neglecting effects of electric size and shape.  These bounds depend only on the relative volume of an object and its material properties.  Shape independent bounds of this kind are necessarily loose and, in general, can only be approached by inverse design procedures in special cases. In the interest of driving optimal design, there is a need to tighten this class of bounds by making them dependent on both material and the shape of the allocated design region.  Shape dependent bounds adapted in this way for maximum radiative heat transfer are given in~\cite{molesky2019fundamental,venkataram2019fundamental}. Shape dependent bounds for maximum absorption are given in~\cite{2019_Molesky_Arxiv} and for photonic design in~\cite{Angeris+etal2019}.

Unlike in photonics, the development of shape dependent bounds is fairly mature in the area of antenna theory, where the task of optimizing many important physical parameters may be cast as convex optimization problems in an unknown current distribution.  Once the operator framework relating source currents to various metrics (\eg{} gain, Q-factor, efficiency) is established, the development of new bounds reduces to the task of manipulating and combining optimization problems of canonical forms.  Because of the convex (or relaxed) nature of such optimization problems~\cite{Boyd+Vandenberghe2004}, their solution is deterministic, many times reducing to the solution of a single eigenvalue problem.  Expressing unknown currents in a finite dimensional basis, the finite dimensional matrix operators governing quantities of interest may be readily calculated using the tools developed for solving integral equation problems~\cite{Gustafsson+etal2016a}, \ie{} using the method of moments~\cite{Harrington1968}.  Problems in antenna theory solved in this way include maximization of directive gain~\cite{Uzsoky+Solymar1956,Jelinek+Capek2017,Gustafsson+Capek2019}, maximization of radiation efficiency~\cite{2018_Shahpari_TAP,Jelinek+Capek2017,Jelinek+etal2018}, minimization of Q-factor (analogous to maximizing bandwidth) \cite{Capek+etal2017b,Gustafsson+Nordebo2013}, and the trade-off between these parameters~\cite{Gustafsson+etal2019, Gustafsson+Capek2019}. 

The purpose of this paper is to transfer this general approach to the field of optics, where the fundamental electromagnetic physics remain the same as in classical antenna theory but the metrics of interest shift from antenna parameters to quantities related to scattering, absorption, extinction, and local field enhancement.  To carry this out, we first develop a framework in which to discuss quantities of interest (\eg{} scattered power, local fields) in terms of operators acting on current densities confined to a region of interest.  This general methodology is shared by previous work~\cite{Miller+etal2016}, though here we use notation and nomenclature based heavily on the numerical solution of integral equations in antenna theory~\cite{Harrington1968} which has been used extensively used for developing bounds in that area~\cite{Gustafsson+etal2016a}.  We then proceed by formulating optimization problems giving rise to shape-specific bounds on absorption, scattering, and extinction cross sections as well as local field enhancement (Purcell's factor) and directive scattering.  We formulate each bound using two sets of constraints governing the enforcement of material constitutive relations.  In a few certain cases, the bounds derived here replicate previous results, however we include them here to highlight the generality and flexibility of the approach taken.  Numerical examples are calculated for each derived bound and serve to demonstrate salient features.

\section{Physical components of problems}
The main goal of this paper is to find bounds on various scattering and absorption metrics by formulating optimization problems which are either convex or can be relaxed to a convex form.  Because many of the components and features are shared between several problems, we begin with an overview of the objectives and constraints used throughout the rest of the paper.  Note that throughout the entirety of the paper, time harmonic steady state is assumed and all quantities are implicitly functions of angular frequency~$\omega$ with $\mathop{\mathrm{exp}}(-\iu\omega t)$ dependence.  The wavenumber~$k = \omega/c_0$ is always that of free space, with~$c_0$ being the speed of light in vacuum. Horizontal axes in figures show free-space wavelength~$\lambda = 2\pi / k$, which is in many cases normalized with respect to the radius~$a$ of the smallest sphere circumscribing the scatterer.

\subsection{Objectives}
The physical quantities investigated in this paper may be represented using linear 
\newcommand{\intsymb}{\zeta}
\begin{equation}
    \intsymb_1 = \int_\reg \boldsymbol{a}^*(\rv) \cdot\boldsymbol{J}(\rv) \diff V
    \label{eq:linear-cont}
\end{equation} 
or sesquilinear (quadratic)
\begin{equation}
    \intsymb_2 = \int_{\reg} \int_\reg \Jv^*(\rv_1) \cdot  \boldsymbol{A}(\rv_1,\rv_2)\cdot\Jv(\rv_2) \diff V_2\diff V_1
    \label{eq:bilinear-cont}
\end{equation}
forms in a current density distribution $\boldsymbol{J}(\rv)$, with all integrations taken over the current's entire spatial support $\reg\subset\mathbb{R}^3$, see Fig.~\ref{fig:radVSscat1}.  Without loss of generality, we assume an appropriate basis $\{\psiv_n (\rv)\}$ for the current density, \eg{}
\begin{equation}
\label{Eq:Current:Expan}
    \Jv(\rv) \approx \sum_{n=1}^N I_n \psiv_n(\rv),
\end{equation}
is chosen in which these forms reduce to linear 
\begin{equation}
    \intsymb_1 \approx \mat{a}^\herm\Jm
\end{equation}
or quadratic
\begin{equation}
    \intsymb_2 \approx \Jm^\herm\mat{A}\Jm
\end{equation}
forms involving vectors and matrices.  For clarity (and compatibility with finite dimensional optimization tools) we only consider the vector/matrix forms except when their equivalence to the forms in~\eqref{eq:linear-cont} and~\eqref{eq:bilinear-cont} yields additional insight. 

Here we briefly describe the physical interpretation and mathematical properties of the vector and matrix operators describing quantities used in optimization problems throughout this paper.  The exact formulation of these operators are not given here but are readily available in the literature~\cite{Harrington1961,Harrington1968,Gustafsson+etal2016a}.  The following discussion assumes that the matrix or vector forms of all operators strictly follow the analytic features (\eg{} definiteness) of their continuous counterparts. 

\subsubsection{Radiated power, absorbed power, and reactance}
\label{Sec:reP}

\begin{figure}[t]
\centering
\includegraphics[]{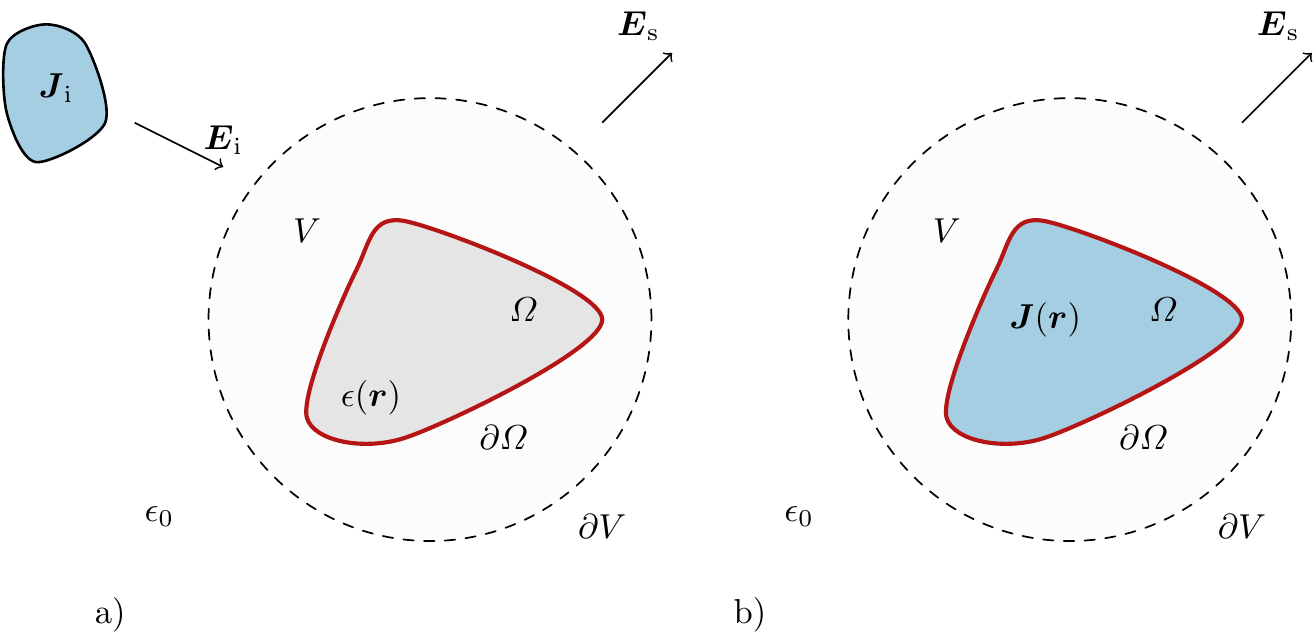}
\caption{An incident electromagnetic field (represented by~$\Ev_{\mrm{i}}$) generating a scattered field (represented by~$\Ev_{\mrm{s}}$). The original problem (a) can be formulated as an equivalent problem (b), where the fields in the presence of (potentially inhomogeneous) materials are replaced by the contrast current density~$\Jv$ impressed in vacuum and radiating the same scattered field $\Ev_{\mrm{s}}$. See App.~\ref{S:PowerBalance} for more details.}
\label{fig:radVSscat1}
\end{figure}

The total cycle-mean power radiated $\pr$ and absorbed $\pa$ by an arbitrary current~$\Jm$ may be cast as quadratic forms~\cite{Gustafsson+etal2016a} 
\begin{equation}
    \pr = \frac{1}{2}\Jm^\herm\Rmr\Jm,
\end{equation}
and
\begin{equation}
\label{Eq:Power:Abs}
    \pa = \frac{1}{2}\Jm^\herm\Rml\Jm,
\end{equation}
respectively. Notice that rigorously the power~$\pr$ refers to a cycle-mean scattered power since the current density~$\Jm$ represents an equivalent contrast source situated in vacuum, see~Fig.~\ref{fig:radVSscat1} and App.~\ref{S:PowerBalance} for more details. As currents should not radiate negative power we have $\pr\geq 0$ for all currents and equivalently $\Rmr \succeq 0$.  Similarly for currents in lossy media we have $\pa\geq 0$ and thus $\Rml \succeq 0$.  Both operators are real symmetric, and their sum~$\pr + \pa$ represents the total cycle mean real power $\psrc$ released by the current $\Jm$, \ie{}
\begin{equation}
    \psrc = \pr+\pa =  \frac{1}{2}\Jm^\herm\left(\Rmr + \Rml\right)\Jm = \frac{1}{2}\Jm^\herm\Rm\Jm.
\end{equation}

Employing spherical wave decomposition, the symmetric positive definite radiation operator $\Rmr$ may be constructed~\cite{Tayli+etal2018} as
\begin{equation}
\label{Eq:SSH}
    \Rmr = \Sm^\herm\Sm,
\end{equation}
where the inner dimension of the matrix $\Sm$ depends on the number of spherical harmonics used.  The number of necessary spherical harmonics, and thus the rank of $\Rmr$, may be approximated using the electrical size of the current support under consideration.  In general, electrically smaller support regions correspond to lower rank.   

The reactance of a current, or equivalently the cycle mean difference in stored magnetic and electric energies, $\wm$ and $\we$, may be calculated via the quadratic form \cite{Harrington+Mautz1971}
\begin{equation}
    2\omega\left(\wm - \we\right) = \frac{1}{2}\Jm^\herm\Xm\Jm.
\end{equation}
The reactance matrix $\Xm$ is real symmetric and, in most cases, indefinite.  Like the operator $\Rm$, the matrix $\Xm$ may be decomposed into two components, one material dependent and one material independent, see Sec.~\ref{Sec:ContrastCurr}.  Together, the matrices $\Rm$ and $\Xm$ constitute the impedance matrix $\Zm = \Rm-\iu\Xm$, see Sec.~\ref{Sec:PowerConst} and App.~\ref{S:PowerBalance}.

%Resonance is defined here in the classical circuit sense of having equal parts stored electric and magnetic energy \cite{Pozar1998}.  Thus a current is self-resonant if $\Jm^\herm\Xm\Jm = 0$.

\subsubsection{Radiation intensity}
\label{Sec:FU}
The scattered field component~$\evh\cdot\boldsymbol{E}_\mathrm{s}(\rv)$ produced by the current $\Jm$ in the $\rvh = \rv/|\rv|$ direction at distance $r\rightarrow\infty$ with polarization~$\evh$ is represented by the linear form
\begin{equation}
\label{Eq:Efar}
    \lim_{r\rightarrow\infty}\evh\cdot \frac{\Ev_\mathrm{s} (\rv)}{\sqrt{\eta_0}}  r\eu^{-\iu k r} = \Fm^\herm\Jm
\end{equation}
such that the corresponding partial radiation intensity is~\cite{Balanis1997}
\begin{equation}
    U(\rvh,\evh) = \frac{1}{2}|\Fm^\herm\Jm|^2 = \frac{1}{2}\Jm^\herm\Fm \Fm^\herm\Jm.
\end{equation}
From the above expression we observe the rank-1 nature of the operator $\Fm\Fm^\herm$, a feature that will enable certain problems to be solved in a computationally efficient manner.  Note that the operator $\Fm\Fm^\herm$ depends directly on the observation direction and polarization.  Like the total absorbed and radiated power operators, the operator  $\Fm\Fm^\herm$ is positive semi-definite and Hermitian symmetric.  

\subsubsection{Radiation enhancement and Purcell's factor}
\label{Sec:PurcellDef}
The cycle mean power radiated by an electric dipole moment~$\boldsymbol{p}$ in the presence of a scatterer is~\cite{Jackson1999}
\begin{equation}
\label{Eq:PFPin}
\pr = - \frac{\omega}{2} \Im \{  \boldsymbol{p}^*\cdot \boldsymbol{E}_\mathrm{s} ( \boldsymbol{r}_p )\} + P_{\rad,p} - \pa,
\end{equation}
where~$\boldsymbol{E}_\mathrm{s} \left( \boldsymbol{r}_p \right)$ is an electric field generated by the scatterer at the point of the dipole,~$\pa$ is the cycle mean power lost within the scatterer~\eqref{Eq:Power:Abs} and
\begin{equation}
\label{Eq:PFPrad0}
P_{\rad,p} = \frac{c_0 k^4}{12\pi \epsilon_0} \left|\boldsymbol{p} \right|^2
\end{equation}
is the cycle mean power radiated by the dipole in free space, with $\epsilon_0$~being permittivity of vacuum (the assumed background medium).

Analogously to the electric far field in~\eqref{Eq:Efar}, the following linear form can be defined
\begin{equation}
\label{Eq:Purcell:Pin}
    - \frac{\omega}{2} \Im \{ \boldsymbol{p}^* \cdot \boldsymbol{E}_\mathrm{s} ( \boldsymbol{r}_0 ) \} = \frac{1}{2} \Re \{ \Jm^\herm \mat{N} \},
\end{equation}
from which a Purcell's factor~\cite{2006_Novotny} can be defined as
\begin{equation}
\label{Eq:Purcell:Def}
F = \frac{\pr}{P_{\rad,p}} = 1 + \dfrac{\Re \{ \Jm^\herm \mat{N} \} - \Jm^\herm \Rml \Jm}{2 P_{\rad,p}}
\end{equation}
which assumes that the back reaction of the dipole to the scatterer may be neglected~\cite{2010_Koenderink_OL, 2015_Krasnok_SR}. This Purcell's ratio characterizes radiation enhancement provided by the scatterer, a metric of primary importance in many areas of applied optics.

Note that unlike in~\eqref{Eq:Purcell:Def}, Purcell's factor might also be defined without the subtraction of the loss term~\cite{2015_Krasnok_SR}. Although in the experimental characterization this might be the only option, evaluating bounds on Purcell's factor in the presence of realistic scatterers without subtracting losses leads to bounds that are too optimistic, as the Purcell's factor may be increased through absorption rather than radiation.

\subsection{Real and reactive power constraints}
\label{Sec:PowerConst}
An incident electric field $\boldsymbol{E}_\mathrm{inc}(\rv)$ may be projected onto the chosen basis in~\eqref{Eq:Current:Expan} via
\begin{equation}
\label{Eq:Vvec}
    \boldsymbol{E}_\mathrm{i}(\rv) \rightarrow \Vm : V_m = \int_\varOmega \boldsymbol{\psi}_m(\rv) \cdot \boldsymbol{E}_\mathrm{i}(\rv) \diff V.
\end{equation}
%When such an incident field takes the form of a plane wave, the propagation direction $\kvh$ and polarization $\evh$ may be denoted via $\boldsymbol{E}_\mathrm{i}(\boldsymbol{\hat{k}},\boldsymbol{\hat{e}}) \rightarrow \Vm(\boldsymbol{\hat{k}},\boldsymbol{\hat{e}})$.
Such an excitation~$\Vm$ uniquely induces a current distribution~$\Jm$ obeying
\begin{equation}
    \Zm \Jm = (\Rm + \iu\Xm )\Jm = \Vm,
    \label{eq:viz}
\end{equation}
where $\Zm$ is the impedance matrix representing the underlying integral operator mapping currents to fields~\cite{Harrington1968}, see App.~\ref{S:PowerBalance} for more details.  For the purpose of formulating general bounds on the behavior of currents existing with a given support, we relax this expression through testing with the current itself, \ie{}
\begin{equation}
    \Jm^\herm\Zm\Jm=\Jm^\herm (\Rm + \iu\Xm)\Jm = \Jm^\herm\Vm, 
    \label{eq:poynting}
\end{equation}
which is an algebraic representation of power conservation~\cite[\S 6.9]{Jackson1999}, see~App.~\ref{S:PowerBalance} for more details. Considering the excitation field~$\Vm$ is given, two power constraints are introduced from~\eqref{eq:poynting} as
\begin{align}
\label{Eq:Pre:Conserv2}
    \Jm^\herm \Rm \Jm - \Re\{\Jm^\herm\Vm\} &= 0, \\
\label{Eq:Pre:Conserv3}     
    \Jm^\herm \Xm \Jm - \Im\{\Jm^\herm\Vm\} &= 0,
\end{align}
representing the conservation of real (cyclic mean) and reactive power, respectively.  The expression in \eqref{Eq:Pre:Conserv2} may be interpreted as a form of the optical theorem relating the total extincted power to that removed from the incident field~\cite[\S 10.11]{Jackson1999}. For plane wave incidence the equivalence to the optical theorem may be further used to relate the extincted power to a forward scattering amplitude. 

As discussed in greater detail in Sec.~\ref{sec:optCurrents}, use of the constraint in~\eqref{Eq:Pre:Conserv2} within an optimization over the current~$\Jm$ amounts to enforcing both loss and radiation properties of an object when illuminated by the incident field~$\Vm$.  For this reason we denote the use of this constraint as using ``prescribed losses'' (indicated throughout the paper by the subscript~$\mathbf{R}$). Similarly, inclusion of the constraint in~\eqref{Eq:Pre:Conserv3} enforces material reactance properties, so together the use of \eqref{Eq:Pre:Conserv2} and \eqref{Eq:Pre:Conserv3} is denoted as using ``prescribed materials'' (indicated throughout the paper by the subscript~$\mathbf{Z}$).

\subsection{Material properties and contrast current}
\label{Sec:ContrastCurr}

Throughout this paper we consider isotropic\footnote{Notice that addition of anisotropy and lossless dielectric background media introduces no difficulties. In such a case, the susceptibility becomes a second-rank tensor $\boldsymbol{\chi}$ and its inversion must be interpreted as a matrix inverse.}, non-magnetic materials with the constitutive relation
\begin{equation}
    \boldsymbol{D} = \epsilon\boldsymbol{E} = \epsilon_0(1+\chi)\boldsymbol{E},
\end{equation}
where~$\epsilon$ is the frequency dependent permittivity and~$\chi$ is the frequency dependent susceptibility.  In the presence of the electric field $\boldsymbol{E}$, the contrast (polarization) current density~$\vec{J}$ may be written in terms of the complex resistivity~$\rho$ as
\begin{equation}
    \boldsymbol{J} = - \iu \omega \epsilon_0\chi \boldsymbol{E} = \rho^{-1}\boldsymbol{E},
    \label{Eq:polCurr}
\end{equation}
where $\rho$ has real and imaginary parts
\begin{equation}
\rho = \rhor+\iu\rhoi 
= \frac{\Im\chi}{\omega\epsilon_0|\chi|^2} + \iu \frac{\Re\chi}{\omega\epsilon_0|\chi|^2}
= \frac{\eta_0\Im\chi}{k|\chi|^2} + \iu \frac{\eta_0\Re\chi}{k|\chi|^2}.
\label{eq:resistivity}
\end{equation}
The real part of resistivity~$\rhor$ is related to a material figure of merit~$\zeta = |\chi|^2 / \Im\chi$ introduced in~\cite{Miller+etal2016,2019_Molesky_Arxiv} as $\rhor / a = \eta_0 / (\zeta ka)$. This last equality can be used in Sec.~\ref{S:MaxOptR} to directly compare the results obtained in this paper with the results obtained in~\cite{2019_Molesky_Arxiv}.

Under the discretization in \eqref{Eq:Current:Expan}, the complex resistivity governs the nature of certain physical quantities and their associated operators.  In general the radiation operator $\Rmr$ does not depend on the resistivity, while the loss operator depends on its real component $\rhor$, \ie{}
\begin{equation}
    \Rm = \Rmr + \Rml.
\end{equation}
The situation is analogous for the reactance operator
\begin{equation}
    \Xm = \Xm_0 + \Xmm.
\end{equation}
For homogeneous bodies, the matrices $\Rml$ and $\Xmm$ scale linearly with the real and imaginary components of the resistivity, respectively.

\subsection{Formulating bounds using optimal currents}
\label{sec:optCurrents}

\begin{figure}[t]%
\centering
\includegraphics[]{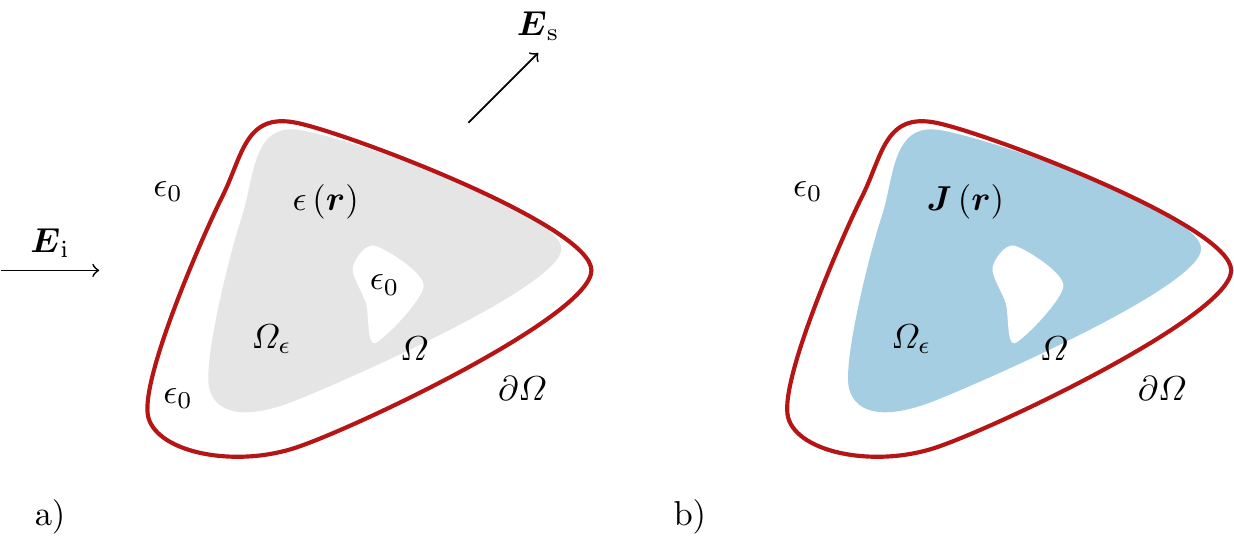}
\caption{Geometry used to determine the maximum absorption, scattering, and extinction of obstacles confined to the region $\reg$ and having permittivity~$\epsilon$. a) The incident $\Ev_{\mrm{i}}$ and $\Ev_{\mrm{s}}$ scattered fields propagate in the background permittivity~$\epsilon_0$ and the obstacle has permittivity $\epsilon$ in the arbitrary subregion $\reg_\epsilon\subseteq\reg$. b) Representation using contrast current density~\eqref{Eq:polCurr} $\Jv$ with support $\supp(\Jv)\subseteq\reg_\epsilon$.}%
\label{fig:scatteringgeo}%
\end{figure}

In the remainder of the paper, we formulate bounds maximizing absorption, scattering, and extinction cross sections, directed scattering, and Purcell's factor for an obstacle with resistivity $\rho=\rhor+\iu\rhoi$ confined to a region~$\reg$, excited by a prescribed incident field.  To do this, we construct and maximize relevant quantities over all possible current density distributions~$\Jv$ supported in~$\reg$, subject to certain constraints and fixed material parameters. These constraints (see~\eqref{Eq:Pre:Conserv2} and~\eqref{Eq:Pre:Conserv3}) enforce conservation of particular quantities (\eg{} real power) but do not mandate that the optimized current be directly excitable by a given incident field.  In this way, the formulated optimization problems yield bounds for all possible structures $\reg_\epsilon \subseteq \reg$ supported within the region $\reg$ consisting of either the structure medium (permittivity $\epsilon$) or the background medium (background permittivity $\epsilon_0$), as shown in Fig.~\ref{fig:scatteringgeo}.  Notice that the exclusion of a given subregion is realized in this paradigm by setting the contrast current density equal zero in that subregion. Because the optimization method is allowed to set currents within any such subregion in $\reg$ to zero, all possible exclusions are automatically considered. This concept of searching for optimal currents has extensively been used in the study of bounds on antenna performance~\cite{Gustafsson+etal2016a}.

As stated previously, we may interpret the application of the real power conservation bound in~\eqref{Eq:Pre:Conserv2} as enforcing only the prescription of material losses $\rhor$. The imaginary (reactive) part of the resistivity $\rhoi$ does not enter into this constraint and may be considered to be chosen freely.  This implies that, under only the constraint of~\eqref{Eq:Pre:Conserv2}, the conservation of reactive power in~\eqref{Eq:Pre:Conserv3} may always be satisfied by a suitable \textit{a posteriori} choice of bulk reactivity $\rhoi$. When both the conservation of real power~\eqref{Eq:Pre:Conserv2} and of reactive power~\eqref{Eq:Pre:Conserv3} are used, an optimization problem with two constraints is formed which is always more restrictive then the former, since now the reactance part of resistivity $\rhoi$ is prescribed prior to the optimization.  

Application of both constraints in~\eqref{Eq:Pre:Conserv2} and~\eqref{Eq:Pre:Conserv3} effectively allows for the calculation of bounds on structures synthesized from fully known and characterized materials (such as gold).  Though looser, application of only the real power bound~\eqref{Eq:Pre:Conserv2}, allows for the examination of hypothetical future materials which may have fixed losses but tunable bulk reactivity.

\section{Maximization of cross sections}
\label{S:CrossSec}

In this section we study upper bounds on the absorption, scattering, and extinction cross sections, denoted as $\Ca$, $\Cs$, and $\Ce$, respectively, achievable by an arbitrary object constructed of material with resistivity~$\rho$ confined to the support~$\reg$. In all cases, a fixed incident field $\Vm$ is assumed which, in the case of cross sections, corresponds to a plane wave~\cite{Bohren+Huffman1983,Kristensson2016}.  Two forms of constraints are considered.  We begin with the simpler of the two, where only the real part $\rhor$ of the resistivity is prescribed (corresponding to constraint~\eqref{Eq:Pre:Conserv2}).  We then extend the problem to include both real and imaginary components of the resistivity (corresponding to both \eqref{Eq:Pre:Conserv2} and \eqref{Eq:Pre:Conserv3}), see Tab.~\ref{Tab1:Problems} for corresponding section numbers for each problem.

\subsection{Prescribed losses}
\label{S:MaxOptR}

\begin{table}[]
    \begin{center}
	\begin{tabular}{cccccc}
	&&\multicolumn{2}{c}{Constraints}& Relative &\\
	Problem & Quantity & \eqref{Eq:Pre:Conserv2} &\eqref{Eq:Pre:Conserv3} &  Complexity &Section \\ \toprule
	maximum absorbed power & \cellcolor{ComplexB}$\paR$ & \cellcolor{ComplexB}$\checkmark$ & \cellcolor{ComplexB} & \cellcolor{ComplexB}$\star\star$ & \cellcolor{ComplexB}\ref{Sec:abs-R} \\
	 & \cellcolor{ComplexC}$\paRX$ & \cellcolor{ComplexC}\checkmark & \cellcolor{ComplexC}\checkmark & \cellcolor{ComplexC}$\star\star\star$ & \cellcolor{ComplexC}\ref{Sec:abs-RX} \\ 
	maximum scattered power & \cellcolor{ComplexB}$\prR$ & \cellcolor{ComplexB}$\checkmark$ & \cellcolor{ComplexB} & \cellcolor{ComplexB}$\star\star$ & \cellcolor{ComplexB}\ref{Sec:scat-R} \\
	 & \cellcolor{ComplexC}$\prRX$ & \cellcolor{ComplexC}\checkmark & \cellcolor{ComplexC}\checkmark & \cellcolor{ComplexC}$\star\star\star$ & \cellcolor{ComplexC}\ref{Sec:scat-RX} \\  
	 maximum extincted power & \cellcolor{ComplexA}$\psrcR$ & \cellcolor{ComplexA}$\checkmark$ & \cellcolor{ComplexA} & \cellcolor{ComplexA}$\star$ & \cellcolor{ComplexA}\ref{Sec:ext-R} \\
	 & \cellcolor{ComplexB}$\psrcRX$ & \cellcolor{ComplexB}\checkmark & \cellcolor{ComplexB}\checkmark & \cellcolor{ComplexB}$\star\star$ & \cellcolor{ComplexB}\ref{Sec:ext-RX}\\ \hline
	 maximum radiation intensity & \cellcolor{ComplexA}$U_\Rm$ & \cellcolor{ComplexA}$\checkmark$ & \cellcolor{ComplexA} & \cellcolor{ComplexA}$\star$ & \cellcolor{ComplexA}\ref{Sec:dir-R} \\
	 & \cellcolor{ComplexB}$U_\Zm$ & \cellcolor{ComplexB}\checkmark & \cellcolor{ComplexB}\checkmark & \cellcolor{ComplexB}$\star\star$ & \cellcolor{ComplexB}\ref{Sec:dir-RX}\\ \hline
	 maximum Purcell's factor & \cellcolor{ComplexB}$F_\Rm$ & \cellcolor{ComplexB}$\checkmark$ & \cellcolor{ComplexB} & \cellcolor{ComplexB}$\star\star$ & \cellcolor{ComplexB}\ref{Sec:pur-R} \\
	 & \cellcolor{ComplexC}$F_\Zm$ & \cellcolor{ComplexC}\checkmark & \cellcolor{ComplexC}\checkmark & \cellcolor{ComplexC}$\star\star\star$ & \cellcolor{ComplexC}\ref{Sec:pur-RX}\\ \bottomrule
	\end{tabular}
    \end{center}	
	\label{Tab1:Problems}
	\caption{Optimization problems solved in the paper.}
\end{table}

\subsubsection{Absorption}
\label{Sec:abs-R}
Maximization of the absorbed power~\eqref{Eq:Power:Abs} with prescribed losses over all possible current densities is determined by the solution to the optimization problem
\begin{equation}
\begin{aligned}
	& \maximize && \pa\\
	& \subto &&  \pa + \pr  = \psrc,
\end{aligned}  
\label{eq:Ca_opt}
\end{equation}
which can, with the help of Sec.~\ref{Sec:reP}, be written as a primal QCQP
\begin{equation}
\begin{aligned}
	& \maximize && \Jm^\herm\Rml\Jm\\
	& \subto &&  \Jm^\herm(\Rml+\Rmr)\Jm - \Re\{\Jm^\herm \Vm\}  = 0.   
\end{aligned}  
\label{eq:Ca_QCQP}
\end{equation}
This QCQP is not convex~\cite{Boyd+Vandenberghe2004} but can be analyzed using the techniques in App.~\ref{APP:Q2CQP}. Particularly, the maximum absorbed power is determined from solution to a dual problem
\begin{equation}
  \paR = \min_{\lagMula > 1}
\frac{\lagMula^2}{8}\Vm^\herm\big((\lagMula-1)\Rml+\lagMula\Rmr\big)^{-1}\Vm,
\label{eq:Pa_dualsol}
\end{equation}
where the condition~$\nu > 1$ follows from the low-rank representation of the matrix~$\Rmr$, see App.~\ref{APP:Q2CQP} for more details. This dual problem can directly be solved using a line search algorithm such as the Newton or bisection algorithms~\cite{Boyd+Vandenberghe2004}. 
Like many of the other problems detailed in later sections, solution to \eqref{eq:Pa_dualsol} can advantageously be formulated in a basis~$\mat{Q}$ which simultaneously diagonalize the operators~$\Rml$ and~$\Rmr$. Vectors of this basis (columns~$\mat{I}_n$ of matrix~$\mat{Q}$) are called radiation modes, see App.~\ref{S:radmodes}, and are the solution to a generalized eigenvalue problem~\cite{Golub+Loan2013}
\begin{equation}
\Rmr \mat{I}_n= \radm_n \Rml \mat{I}_n.
\label{eq:radmodes}
\end{equation}
The radiation matrix~$\Rmr$ is known to be of low rank, and can advantageously be decomposed as~$\Rmr=\Sm^\herm\Sm$, see~\cite{Tayli+etal2018}. The loss matrix~$\Rml$ is a sparse full rank positive definite matrix and can be decomposed via a Cholesky decompostion~\cite{Golub+Loan2013} as~$\Rml=\Rmlf^\herm\Rmlf$. Substituting into~\eqref{eq:radmodes} and using a singular value decomposition~(SVD)~\cite{Golub+Loan2013}~$\Sm\Rmlf^{-1} = \Um_1\Sigmam\Um_2^\herm$, where matrices~$\Um_1,\Um_2$ are unitary and matrix~$\Sigmam$ is diagonal, it can be shown that $\radm_n = \Sigma_{nn}^2$ and $\mat{Q} = \Rmlf^{-1} \Um_2$.

Assuming $\Jm = \mat{Q} {\tilde\Jm}$, the minimization problem~\eqref{eq:Pa_dualsol}, normalized to incident power flux~$S_0$, can be rewritten as
\begin{equation}
\CaR = \min_{\lagMula > 1} \frac{\lagMula^2}{8 S_0} \sum_{n=1}^N \frac{ \big| {\tilde V}_n \big|^2 } { \lagMula (1 + \radm_n)-1} 
= \min_{\lagMula > 1} \frac{\lagMula^2}{4k^2} \sum_{n=1}^N \frac{ \radm_n |\at_n|^2}{ \lagMula ( 1+\radm_n) - 1},
\label{eq:Ca_R}
\end{equation}
where~${\tilde \Vm} = \mat{Q}^\herm \Vm$. The second equality in~\eqref{eq:Ca_R} assumes that the excitation vector has been composed of spherical waves as~$\Vm = \Sm^\herm \am/(k\sqrt{\eta_0})$, where $\am$ are dimensionless spherical wave expansion coefficients of an incident plane wave with unit amplitude, see App.~\ref{S:radmodes}. A projection~$\amt = \Um_1^\herm \am$ has also been defined to ease the notation.

As shown in~App.~\ref{APP:Q2CQP}, the form in~\eqref{eq:Ca_R} (as compared to~\eqref{eq:Pa_dualsol}) allows for simple analytical evaluation of the first and second derivative with respect to the Lagrange multiplier~$\lagMula$ being well prepared for a minimization via Newton's algorithm. Furthermore, the summation involved in~\eqref{eq:Ca_R} converges quickly in realistic scenarios, since for small electrical sizes ($ka \leq 1$), the eigenvalues~$\radm_n$ decay rapidly in $n$ and only a few terms are needed, see~App.~\ref{S:radmodes}. Notice that a similar decomposition was implicitly used in~\cite[Eq. 4, 5]{2019_Molesky_Arxiv} to solve a related optimization problem.

\subsubsection{Scattering}
\label{Sec:scat-R}

Maximization of the scattered power is determined by interchanging $\Rml$ and $\Rmr$ in~\eqref{eq:Ca_QCQP} and~\eqref{eq:Pa_dualsol} which gives
\begin{equation}
\begin{aligned}
	& \maximize && \Jm^\herm\Rmr\Jm\\
	& \subto &&  \Jm^\herm(\Rml+\Rmr)\Jm - \Re\{\Jm^\herm \Vm\}  = 0
\end{aligned}  
\label{eq:Ps_QCQP}
\end{equation}
with the dual problem
\begin{equation}
  \prR
=\min_{\lagMula > \lagMula_1}
\frac{\lagMula^2}{8}\Vm^\herm\big(\lagMula\Rml+(\lagMula-1)\Rmr\big)^{-1}\Vm,
\label{eq:Ps_dualsol}
\end{equation}
where $\lagMula_1=\radm_1/(1+\radm_1)$ with $\radm_1$ the largest eigenvalue~\eqref{eq:radmodes} for the radiation modes.  Observe that this value of $\lagMula_1$ coincides with the maximum radiation efficiency for antennas restricted to the region $\reg$~\cite{Gustafsson+etal2019}.

Analogously to the bound on the absorption cross section in~\eqref{eq:Ca_R}, the bound on the scattering cross section can be written in terms of radiation modes as
\begin{equation}
\CsR = \min_{\lagMula > \lagMula_1} \frac{\lagMula^2}{8 S_0} \sum_{n=1}^N \frac{ \big| {\tilde V}_n \big|^2 } { \lagMula \left( 1 + \radm_n \right) - \radm_n } 
= \min_{\lagMula > \lagMula_1} \frac{\lagMula^2}{4k^2} \sum_{n=1}^N \frac{ \radm_n | {\at}_n|^2 } { \lagMula \left( 1 + \radm_n \right) - \radm_n }.
\label{eq:Cs_R}
\end{equation}

\subsubsection{Extinction}
\label{Sec:ext-R}
The maximum extincted power is determined from maximization of the total cycle mean power~$\psrc$. In this case, the power balance in~\eqref{eq:poynting} can be used to greatly reduce the complexity of the problem into 
\begin{equation}
\begin{aligned}
	& \maximize && \Re\{\Vm^\herm\Jm\}\\
	& \subto &&  \Jm^\herm(\Rml+\Rmr)\Jm - \Re\{\Jm^\herm \Vm\}  = 0    
\end{aligned}  
\label{eq:Pe_QCLP}
\end{equation}
with the explicit solution 
\begin{equation}
  \psrcR = \frac{1}{2}\Vm^\herm(\Rml+\Rmr)^{-1}\Vm = \frac{1}{2}\Vm^\herm\Rm^{-1}\Vm.
\label{eq:Pe_dualsol}
\end{equation}
This expression is also recognized from bounds on the maximum effective area $\Aeff$ of antennas restricted to region~$\reg$~\cite{Gustafsson+Capek2019}.

Following the same methodology as applied to absorption and scattering, the maximum extinction cross section~\eqref{eq:Pe_dualsol} can be written in terms of radiation modes as
\begin{equation}
\CeR = \dfrac{1}{2S_0} \sum_{n=1}^N \frac{ \big| {\tilde V}_n \big|^2}{ 1 + \radm_n} 
=  \frac{1}{k^2}\sum_{n=1}^N \frac{ \radm_n |\at_n|^2}{1 + \radm_n} .
\label{eq:Ce_R}
\end{equation}

\subsubsection{Electrically small scatterers}
\label{S:SmallSize}

Although the bounds in~\eqref{eq:Ca_R},~\eqref{eq:Cs_R}, and~\eqref{eq:Ce_R} are easily determined for an arbitrary shaped geometry, their explicit approximations depending only on resistivity $\rhor$, volume $\vol$, and free-space wavenumber $k$ are of great interest~\cite{Miller+etal2016}. Such approximation is possible in the limit of small electric sizes~$k a = 2 \pi a / \lambda \ll 1$. In this case it can be assumed that the summations in~\eqref{eq:Ca_R}, \eqref{eq:Cs_R} and~\eqref{eq:Ce_R} are dominated by the first term. If only this dominant term ($n = 1$) is kept, an analytical solution exists in all three cases and reads

\begin{equation}
\CaR \stackrel{ka\ll 1}{\approx}
\frac{\radm_1|\at_1|^2} {k^2( 1+\radm_1)^2} 
= \frac{\eta_0 V }{ \rhor \left( 1 + \frac{ \eta_0 k^2 V }{ 6 \pi \rhor} \right)^2 }
\leq 
\begin{cases}
\eta_0 \vol/\rhor=\Carho\\
3\pi/(2k^2)=\Cadip
\end{cases}
\label{eq:Ca_Rlow}
\end{equation}

\begin{equation}
\CsR \stackrel{ka\ll 1}{\approx} 
\frac{\radm_1^2|\at_1|^2}{k^2(1+\radm_1)^2} 
= \frac{k^2}{6\pi} \frac{\left( \frac{\eta_0  V }{\rhor}\right)^2}{\left( 1 + \frac{ \eta_0 k^2 V}{6\pi\rhor}\right)^2 }   
\leq 
\begin{cases}
k^2\eta_0^2 \vol^2 / (6\pi\rhor^2)\\
\eta_0 \vol / (4\rhor)=\Csrho\\
6\pi / k^2=\Csdip\\
\end{cases}
\label{eq:Cs_Rlow}
\end{equation}

\begin{equation}
\CeR \stackrel{ka\ll 1}{\approx} 
\frac{\radm_1 |\at_1|^2}{k^2(1 + \radm_1)}  
= \frac{\eta_0 V }{\rhor \left( 1 + \frac{ \eta_0 k^2 V }{ 6 \pi \rhor} \right)}
 \leq 
\begin{cases}
\eta_0 \vol / \rhor\\
6\pi / k^2
\end{cases}
\label{eq:Ce_Rlow}
\end{equation}
where the dominant radiation mode $\radm_1 \approx \eta_0 k^2 V/(6 \pi\rhor)$ and the projection of the plane wave on a dipole mode $|\at_1|^2 = 6\pi$ for $ka\ll 1$ have been used, see App.~\ref{S:radmodes}.

Relation~\eqref{eq:Ca_Rlow} shows that the absorption cross section is, in the limit of small electric size, bounded by $\Carho$ introduced in~\cite{Miller+etal2016} and by the dipole bound $\Cadip$~\cite{Tretyakov2014}. The material-only bound is valid for all electrical sizes~\cite{Miller+etal2016} and is obtained by neglecting the radiation term $\Rmr$ in~\eqref{eq:Pa_dualsol}. The dipole bound is restricted to dipole interaction and can be obtained from the directivity, $D$, and gain $G$, of a lossless Hertzian dipole $D = G = 3/2$ having effective area $\lambda^2 G/(4\pi)=3\pi/(2 k^2)$~\cite{Silver1949}. These two bounds intersect at $k=\sqrt{3\pi\rhor/(2\eta_0 V)}$.
The bound $\CaR$ has a maximum for $\radm_1 = 1$ but decreases monotonically with increasing wavenumber $k$ for electrically small objects. The single term (dipole) approximation is not valid as $\rhor\to 0$ since the dipole contribution to the absorption is negligible and therefore it is necessary to include higher order terms.

Results for bounds on scattering cross section~\eqref{eq:Cs_Rlow} show three distinct regions at small electrical sizes as compared with the two for the absorption cross section~\eqref{eq:Ca_Rlow}. The scattering cross section first increases with $k^2$ reaching its maximum $\Csrho$ for the wavenumber
\begin{equation}
  k_{\mrm{s}}=\sqrt{\frac{6\pi\rhor}{\eta_0 V}}
\label{eq:kmaxScatt}
\end{equation}
after which it deceases with $k^{-2}$. Small self-resonant objects are often minimum scattering~\cite{Montgomery+Dicke+Purcell1948,Kahn+Kurss1965} and hence $\Ca=\Cs$ which implies
$\radm_1=1$ which is recognized as~\eqref{eq:kmaxScatt} and from the maximum of~\eqref{eq:Cs_Rlow}.  
  
The small electric size limit of the bound on the extinction cross section~\eqref{eq:Ce_Rlow} is similar to the two previous cases. 

\subsubsection{Numerical examples}
\begin{figure}[th]%
{\centering
\includegraphics[width=\columnwidth]{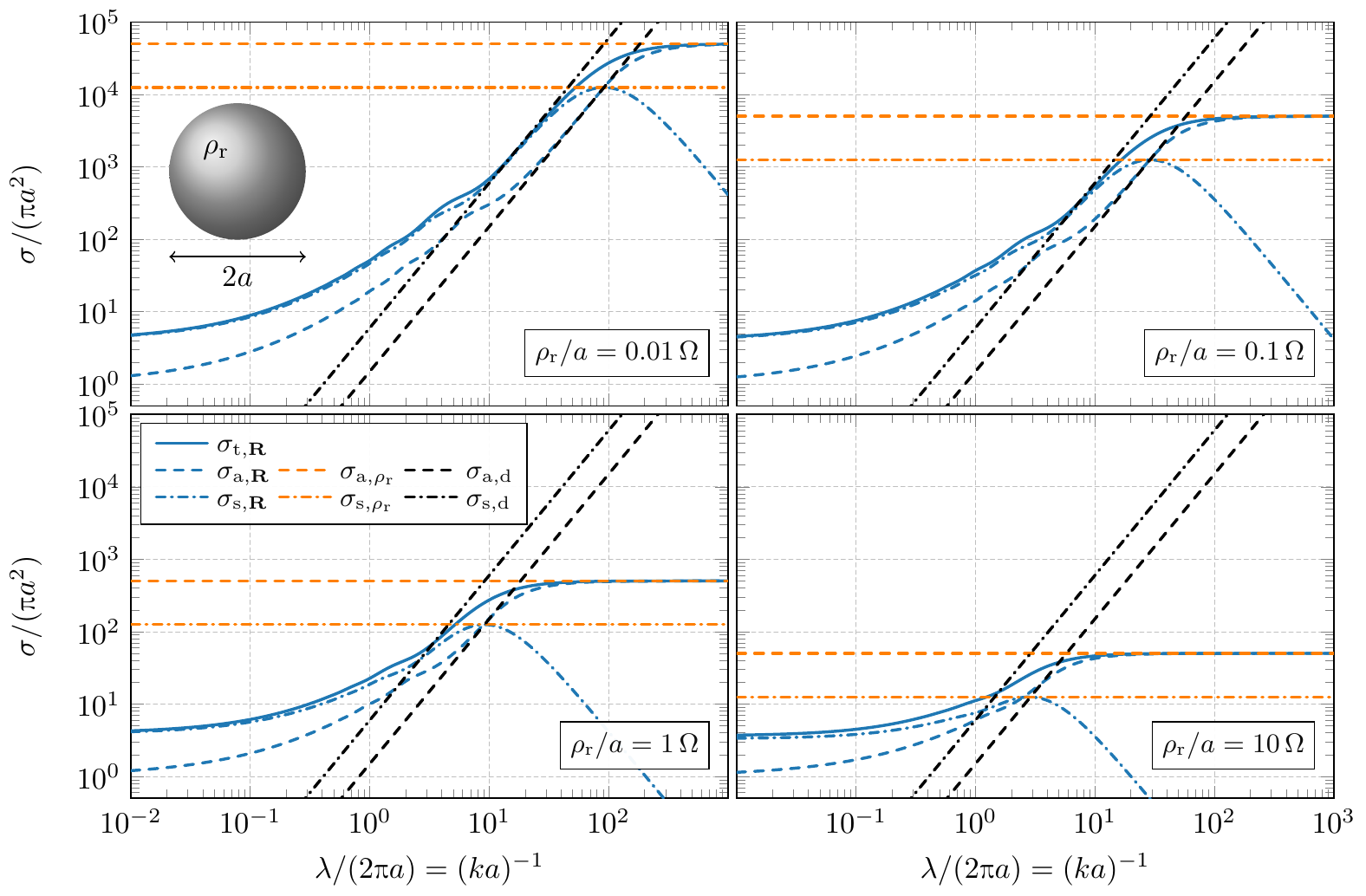}
\par}
\caption{
Comparison between bounds on absorption~\eqref{eq:Ca_R}, scattering~\eqref{eq:Cs_R}, and extinction~\eqref{eq:Ce_R} cross sections for obstacles with resistivities~\eqref{eq:resistivity} $\rhor/a\in \{0.01,0.1,1,10\}\unit{\Omega}$ circumscribed by a sphere with radius $a$. The bounds are compared with the material-only $\sigma_{\ast,\rhor}$ and dipole $\sigma_{\ast,\mrm{d}}$ bounds from the electrically small approximations~\eqref{eq:Ca_Rlow} and~\eqref{eq:Cs_Rlow}.}%
\label{fig:spherebound}%
\end{figure}

Bounds on absorption, scattering, and extinction cross sections are depicted in Fig.~\ref{fig:spherebound} for obstacles supported in a spherical region with radius $a$ and resistivities $\rhor/a\in\{0.01,0.1,1,10\}\unit{\Omega}$. The bounds are computed using an analytical prescription for radiation modes of a sphere~\eqref{eq:radmodes_sph}, normalized with the geometrical cross section $\Across=\pi a^2$, and shown for electric sizes $10^{-3}\leq ka\leq 10^2$. Compared with the small-size approximations~from Sec.~\ref{S:SmallSize} (dashed and dotted lines) it is observed that $\CaR$ follows $\Carho$~\cite{Miller+etal2016} up to $k=k_{\mrm{s}}/2$, \cf~\eqref{eq:kmaxScatt}, where the small size approximations~\eqref{eq:Ca_Rlow} intersect. Then the bound follows the dipole approximation until the structure starts to support higher order modes. The maximum of $\CsR$ is reached at $k=k_{\mrm{s}}$ and is well described by the small size approximation~\eqref{eq:Cs_Rlow}. The dipole approximations are most useful for low-loss cases where they approximate the bounds over a large range of frequencies. The approximate onset of different radiation modes can be found from the condition $\radm_n\approx 1$, with the radiation modes for the sphere in~\eqref{eq:radmodes_sph} depicted Fig.~\ref{fig:Sphradmodes}. The first mode reaches $\radm_1\approx 1$ at $ka\approx k_{\mrm{s}}a= \sqrt{9\rhor/(2\eta_0 a)}\approx \{0.01,0.03,0.11,0.35\}$ and dominates the cross sections until the onset of the second mode ($\radm_2\approx 1$) at $ka\approx \sqrt[4]{18\rhor/(\eta_0 a)}\approx \{0.15,0.26,0.47,0.83\}$, where a $\sqrt[4]{\rhor}$ scaling in $\rhor$ for the second mode is also seen.

The bounds increase monotonically with decreasing resistivity $\rhor$ and are unbounded in the limit $\rhor\to 0$. This increase is, however, slow for $k>k_{\mrm{s}}$ and negligible in the limit of electrically large structures where $\CaR$ approaches the geometrical cross section $\Across$ and $\CsR,\CeR$ approach $4\Across$.  Notice that the bound~$\CaR$ is similar to that derived in~\cite{2019_Molesky_Arxiv}. The results shown in Fig.~\ref{fig:spherebound} for~$\CaR$ can therefore be compared with those in Fig.~2 from~\cite{2019_Molesky_Arxiv} taking into account the relation~$\rhor / a = \eta_0 / (\zeta ka)$.

\begin{figure}
\centering
\includegraphics[]{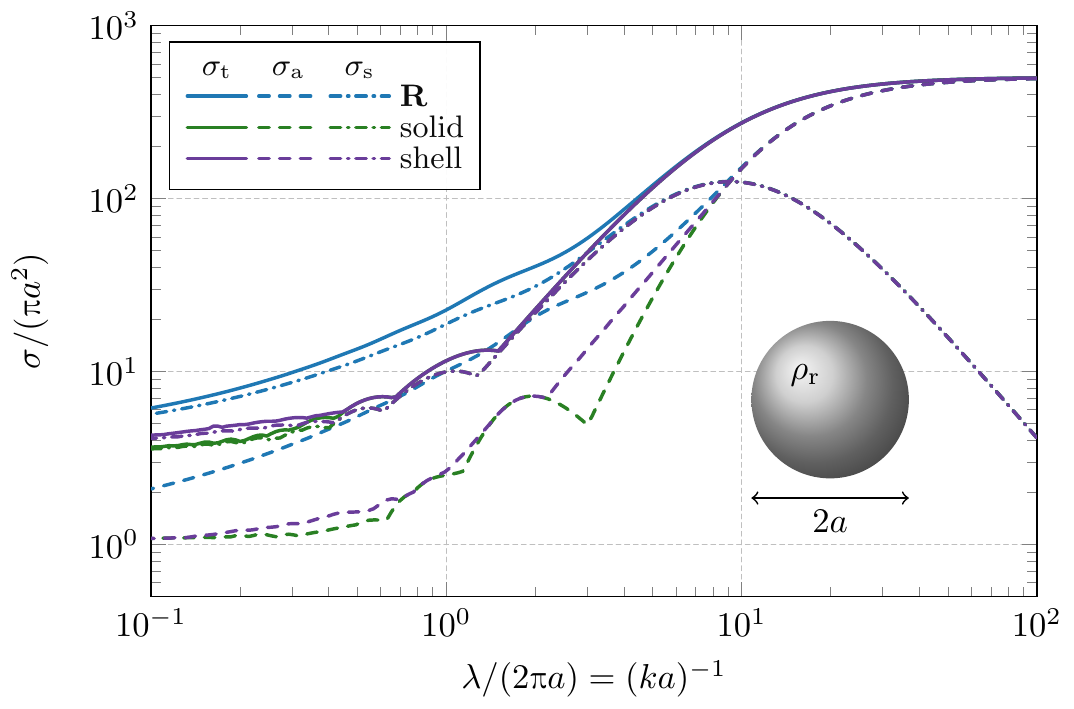}
\caption{Bounds on absorption~\eqref{eq:Ca_R}, scattering~\eqref{eq:Cs_R}, and extinction~\eqref{eq:Ce_R} cross sections for spherical regions compared with realized cross sections of solid spheres and spherical shells homogeneously filled with resistivity~\eqref{eq:resistivity} $\rhor/a= 1\unit{\Omega}$ and optimized reactance $\rhoi$ and shell thickness.}%
\label{fig:BoundsphereCompOptRhoi}%
\end{figure}

Tightness of the bounds is investigated in Fig.~\ref{fig:BoundsphereCompOptRhoi}, where the realized cross sections for solid spheres and spherical shells are compared with the bounds. The imaginary part $\rhoi=\Im\{\rho\}$ and shell thickness are swept numerically for each electric size and the maximum realized cross sections are depicted in Fig.~\ref{fig:BoundsphereCompOptRhoi}. Solid spheres reach the bounds for small electric sizes $ka\leq 0.1$ but are slightly below the bounds for larger sizes. Optimization over the layer thickness increases the cross sections slightly but does not reach the bounds. 

\begin{figure}[t]
\centering
\includegraphics[]{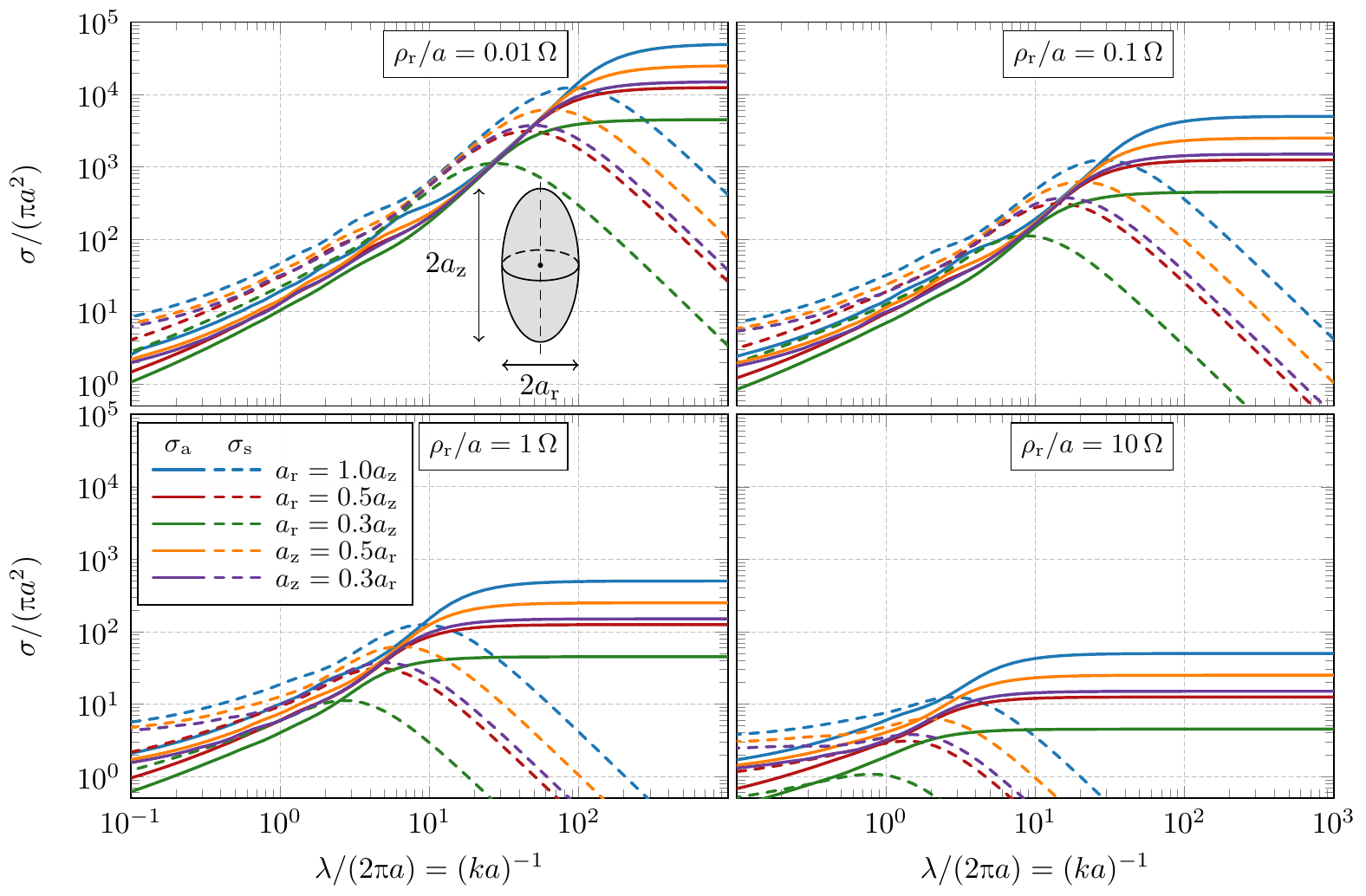}
\caption{Comparison between bounds on absorption~\eqref{eq:Ca_R} and scattering~\eqref{eq:Cs_R} cross sections for spheroids with resistivity~\eqref{eq:resistivity} $\rhor/a\in\{0.01,0.1,1,10\}\unit{\Omega}$ as function of the electrical size.}
\label{fig:spheroidbound}
\end{figure}
Bounds on absorption and scattering cross sections for obstacles circumscribed by spheroidal regions with semi axes $a_{\mrm{r}}$ and $a_{\mrm{z}}$ and an incident plane wave from the $\zvh$-direction are depicted in Fig.~\ref{fig:spheroidbound}. The overall behavior of the bounds for circumscribing spheres in Fig.~\ref{fig:spherebound} and spheroids are similar. A decrease for electrically small spheroid as compared with the sphere $a_{\mrm{r}}=a_{\mrm{z}}$ is explained by the reduced volume, see~\eqref{eq:Ca_Rlow} and~\eqref{eq:Cs_Rlow}. 

\subsection{Prescribed materials}
\label{S:MaxOptXR}
Here we adapt the physical bounds in Sec.~\ref{S:MaxOptR} to a stricter form where both real and imaginary components of the resistivity $\rho$ are fixed. %on the cross sections for obstacles supported in a region $\reg_1\subset\reg$ and made of a material with given permittivity can be determined using the relaxation~\eqref{eq:Zconstr}. The bounds are formulated for a given region $\reg$ and permittivity $\epsilon(\rv)$ or equivalently resistivity $\rho(\rv)$ using~\eqref{eq:resistivity}. The relaxation~\eqref{eq:Zconstr} is valid for all obstacles with a resistivity $\rho_1(\rv)=1_{\reg_1}(\rv)\rho_(\rv)$, where $1_{\reg_1}(\rv)$ is the characteristic function (indicator function) for the region $\reg_1\subset\reg$, \ie $1_{\reg_1}(\rv)=1$ if $\rv\in\reg_1$ and $0$ otherwise.

\subsubsection{Absorption}
\label{Sec:abs-RX}
Adding the reactance condition~\eqref{Eq:Pre:Conserv3} to the optimization problem in \eqref{eq:Ca_QCQP} for maximizing the absorbed power, we obtain
\begin{equation}
\begin{aligned}
	& \maximize && \Jm^\herm\Rml\Jm\\
	& \subto &&  \Jm^\herm\Rm\Jm - \Re\{\Jm^\herm \Vm\}=0   \\
	&  &&  \Jm^\herm\Xm\Jm - \Im\{\Jm^\herm \Vm\}  = 0   
\end{aligned}  
\label{eq:MaxAbsRX}
\end{equation}
and the dual problem (see App.~\ref{APP:Q2CQP})
\begin{equation}
\paRX=\min_{(\lagMula,\lagMulb)\in \Ds}\frac{\lagMula^2+\lagMulb^2}{8}\Vm^\herm\big(-\Rml+\lagMula\Rm+\lagMulb\Xm\big)^{-1}\Vm,
\label{eq:MaxAbsRXdual}
\end{equation}
where $\Ds = \{(\lagMula,\lagMulb):~ -\Rml+\lagMula\Rm+\lagMulb\Xm \succeq\Om\}$.

\subsubsection{Scattering}
\label{Sec:scat-RX}

The scattering cross section is similarly obtained by interchanging $\Rmr$ and $\Rml$ giving the primal problem
\begin{equation}
\begin{aligned}
	& \maximize && \Jm^\herm\Rmr\Jm\\
	& \subto &&  \Jm^\herm\Rm\Jm - \Re\{\Jm^\herm \Vm\}=0   \\
	&  &&  \Jm^\herm\Xm\Jm - \Im\{\Jm^\herm \Vm\}  = 0   
\end{aligned}  
\label{eq:MaxScattRX}
\end{equation}
and the dual problem
\begin{equation}
\prRX=\min_{(\lagMula,\lagMulb)\in \Ds}\frac{\lagMula^2+\lagMulb^2}{8}\Vm^\herm\big(-\Rmr+\lagMula\Rm+\lagMulb\Xm\big)^{-1}\Vm,
\label{eq:MaxScattRXdual}
\end{equation}
where $\Ds = \{(\lagMula,\lagMulb):~ -\Rmr+\lagMula\Rm+\lagMulb\Xm \succeq\Om\}$.

\subsubsection{Extinction}
\label{Sec:ext-RX}
For the extinction cross section, the optimization problem with both constraints is
\begin{equation}
\begin{aligned}
	& \maximize && \Re\{\Jm^\herm \Vm\}\\
	& \subto &&  \Jm^\herm\Rm\Jm - \Re\{\Jm^\herm \Vm\}=0   \\
	&  &&  \Jm^\herm\Xm\Jm - \Im\{\Jm^\herm \Vm\} = 0   
\end{aligned}  
\label{eq:MaxExtRX}
\end{equation}
yielding the dual function
\begin{equation}
%  \inf_{\Jm}  L(\Jm,\lagMula,\lagMulb)
\psrcRX = \min_{(\lagMula,\lagMulb)\in \Ds}
=\frac{(1+\lagMula)^2+\lagMulb^2}{8}\Vm^\herm(\lagMula\Rm+\lagMulb\Xm)^{-1}\Vm,
%=\frac{(1+\lagMula)^2+\lagMula^2\lagMulb_1^2}{4\lagMula}\Vm^\herm(\Rm+\lagMulb_1\Xm)^{-1}\Vm,
\label{eq:MaxExtRXdual}
\end{equation}
where $\Ds = \{(\lagMula,\lagMulb):~ \lagMula\Rm+\lagMulb\Xm\succeq\Om\}$. Substituting $\lagMulb = \lagMula\lagMulb_1$ and carrying out the minimization over $\lagMula$ (now entirely outside of the matrix inversion) gives $\lagMula=\pm 1/\sqrt{1+\lagMulb_1^2}$, which simplifies \eqref{eq:MaxExtRXdual} to
% \begin{equation}
%   \frac{(1+\lagMula)^2+\lagMula^2\lagMulb_1^2}{4\lagMula}
%   =\frac{1+2\lagMula+\lagMula^2(\lagMulb_1^2+1)}{4\lagMula}
%   =\frac{1+\lagMula}{2\lagMula}
%   =\frac{1+1/\lagMula}{2}
%   =\frac{1+\sqrt{1+\lagMula_1^2}}{2}
% \label{eq:}
% \end{equation}
% that simplifies the Lagrangian to  
\begin{equation}
%  \inf_{\Jm}  L(\Jm,\lagMula,\lagMulb)
%=\frac{(1+\lagMula}{\lagMula}\Vm^\herm(\Rm+\lagMulb_1\Xm)^{-1}\Vm
\psrcRX = \min_{+,-}\min_{\lagMulb_1 \in \Ds_{1\pm}} \frac{1\pm\sqrt{1+\lagMulb_1^2}}{4}\Vm^\herm(\Rm+\lagMulb_1\Xm)^{-1}\Vm.
\label{eq:CeZdual}
\end{equation}

Since the optimization problem~\eqref{eq:CeZdual} only involves two matrices, it can still be, analogously to problems in Sec.~\ref{S:MaxOptR}, diagonalized using a basis~$\mat{Q}$ with basis vectors (columns $\mat{I}_n$ of matrix~$\mat{Q}$) generated by an eigenvalue problem
\begin{equation}
  \Xm\Jm_n = \eigv_n\Rm\Jm_n.
\label{eq:cm}
\end{equation} 
These eigenmodes resemble characteristic modes~\cite{Harrington+Mautz1971}, though here, unlike the classical formulation for characteristic modes on perfectly conducting bodies, the operator $\Rm$ contains terms related to both loss and radiation and thus alters the modal set's orthogonality properties \cite{Harrington+etal1972}.

Normalizing~$\mat{Q}^\herm \Rm \mat{Q}$ to be an identity matrix, $\mat{Q}^\herm \Xm \mat{Q}$ is a diagonal matrix of eigennumbers~$\eigv_n $ and the dual formulation~\eqref{eq:CeZdual} can be rewritten as
\begin{equation}
    \psrcRX = \min_{+,-}\min_{\lagMulb_1 \in \Ds_{1\pm}} \frac{1\pm\sqrt{1+\lagMulb_1^2}}{4} \sum_{n=1}^N \frac{|\tilde{V}_n|^2}{1+\lagMulb_1\lambda_n}
    \label{eq:psrcRXDiag}
\end{equation}
where~$\tilde{\mat{V}}=\mat{Q}^{\herm}\Vm$ and the $+$ sign corresponds to the domain~$\Ds_{1+}$ while the $–$~sign corresponds to the domain~$\Ds_{1-}$. These domains depend on the definiteness of the matrix~$\Xm$ and are defined as
\begin{equation}
    \left\{\hspace{-1ex}\begin{tabular}{lll}
         $\Ds_{1+} =[-( \max \lambda_n )^{-1}, -( \min \lambda_n )^{-1}]$, & $\Ds_{1-}=\emptyset$, & $\Xm$ indefinite\\
         $\Ds_{1+} =[-( \max \lambda_n )^{-1},\infty]$, & $\Ds_{1-} =[-\infty,-( \min \lambda_n )^{-1}] $   & $\Xm\succeq 0$\\ 
    %     \\
         $\Ds_{1+} =[-\infty,-( \min \lambda_n )^{-1}]$, &
         $\Ds_{1-} =[-( \max \lambda_n )^{-1},\infty]$ &
         $\Xm \preceq 0$ \\
    \end{tabular}\right.
    \label{eq:CeZrange}
\end{equation}
In \eqref{eq:psrcRXDiag}, the inner minimization is performed first for the multiplier~$\lagMulb_1$ in the domain~$\Ds_{1+}$ and~$\Ds_{1-}$ separately and from those two results, the minimum is selected by the outer minimization.

\subsubsection{Numerical examples}
\begin{figure}[thb]
\centering
\includegraphics[width=0.95\columnwidth]{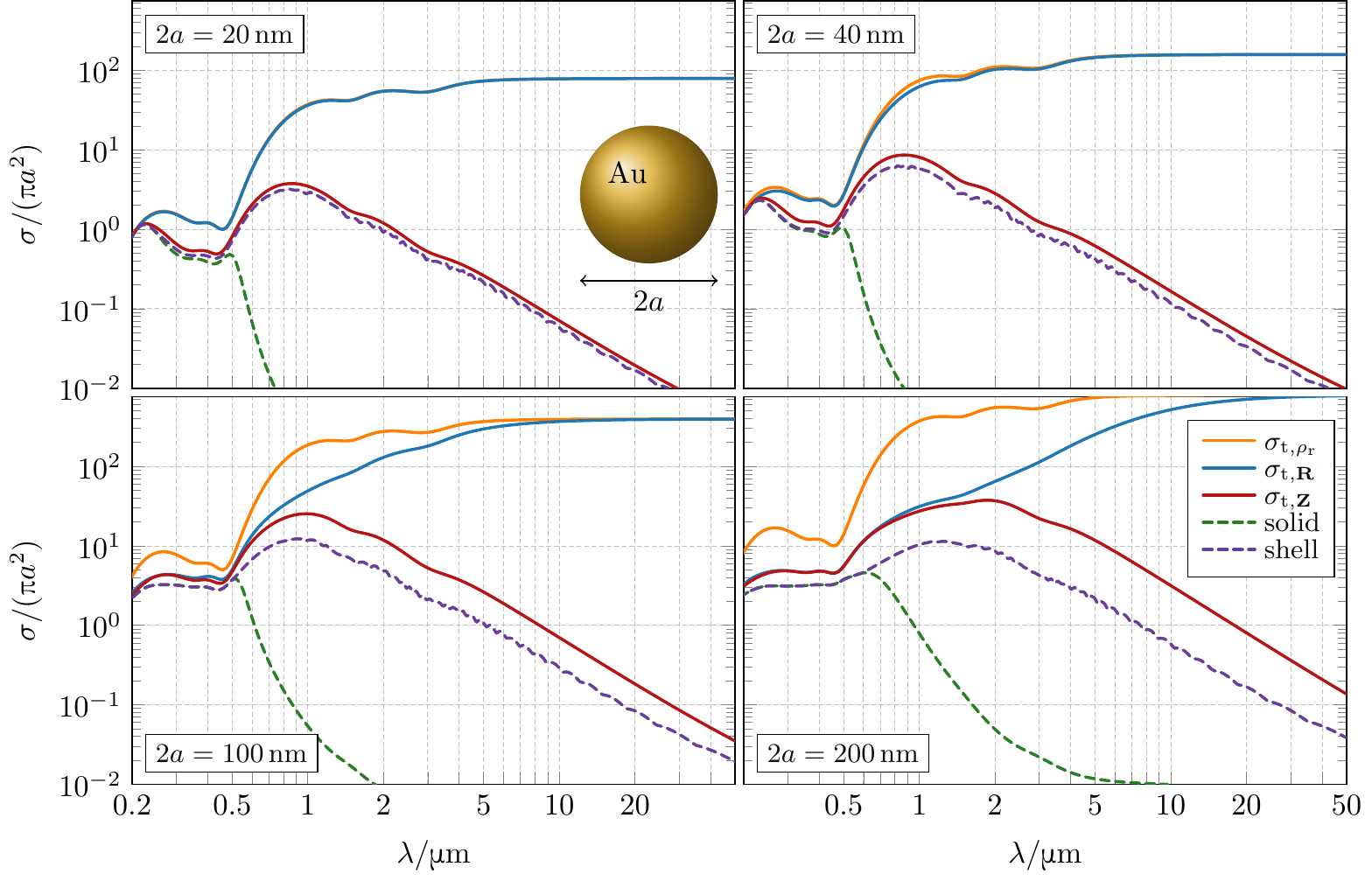}
\caption{Bounds~\eqref{eq:Ce_R},~\eqref{eq:Ce_Rlow}, and~\eqref{eq:psrcRXDiag} on the extinction cross sections for gold obstacles confined in a spherical region with radii $a\in\{10,20,50,100\}\unit{nm}$. The bounds are compared with the extinction cross section for a homogeneous gold sphere and an optimized gold spherical shell.}
\label{fig:Aubounds}
\end{figure}

Bounds on the extinction cross section are depicted in Fig.~\ref{fig:Aubounds} for obstacles composed of gold (Au), see App.~\ref{S:MaterialModels}, and circumscribed by a sphere with radius $a\in\{10,20,50,100\}\unit{nm}$. The bounds~\eqref{eq:Ce_Rlow}, \eqref{eq:Ce_R}, and~\eqref{eq:psrcRXDiag} are compared with the realized extinction cross sections of solid gold spheres and spherical gold shells where the shell thickness is optimized to maximize the extinction cross section $\Ce$. For all radii it is clear that inclusion of the reactance constraint~\eqref{Eq:Pre:Conserv3} has a large effect for longer wavelengths or, more accurately, electrically small sizes ($ka = 2 \pi a / \lambda \leq 1$). The effect diminishes as the electric size increases and is negligible for $ka\geq 1$, \cf~\cite{Gustafsson+Capek2019}. 

Starting with radius $a=10\unit{nm}$, it is observed that $\Cerho\approx\CeR$. Inclusion of the reactance constraint~\eqref{Eq:Pre:Conserv3} reduces the bound on $\Ce$ by orders of magnitude for wavelengths~$\lambda\geq 1\unit{\micro m}$. The differences are smaller for shorter wavelengths and minuscule for~$\lambda\approx 0.2\unit{\micro m}$ and $\lambda\approx 0.5 \unit{\micro m}$. The longer wavelength $0.5\unit{\micro m}$ corresponds to the dipole plasmonic resonance~\cite{Maier2007,Bohren+Huffman1983} which occurs for a relative permittivity with a real part close to $-2$ (Fröhlich condition), see Fig.~\ref{fig:Aubounds}, but the high losses weaken this effect. The extinction cross section for homogeneous spheres is also close to the $\CeZ$ bound for $\lambda\in[0.2,0.5]\unit{\micro m}$ but far from the bound for longer wavelengths. The performance improves for spherical shells with shell thickness optimized for maximum $\Ce$ which are seen to follow the $\CeZ$ bound for the considered wavelengths. This is partly explained by the added degree of freedom from the shell thickness which is used to tune the plasmonic resonance at each wavelength. The conclusions are similar for the larger radii but $\Cerho$ and $\CeR$ starts to differ and the difference is large for $a\geq 50\unit{nm}$. Performance of the solid spheres and spherical shells relative to the bounds worsen with increasing radius. 

\begin{figure}[t]
\centering
\includegraphics[]{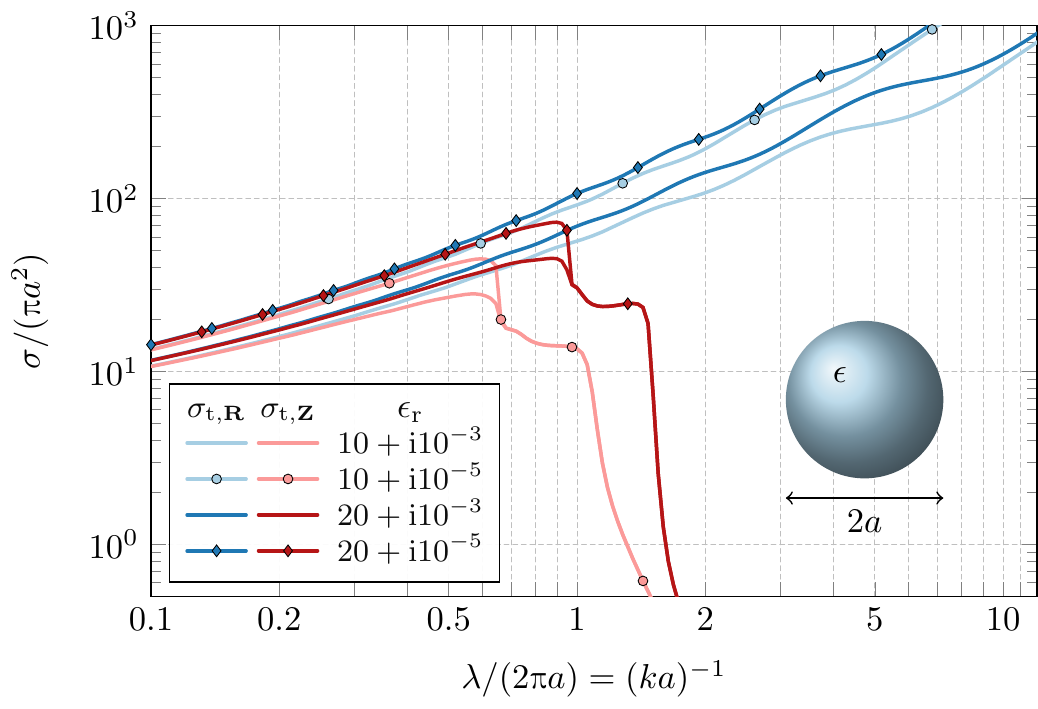}
\caption{Upper bounds~\eqref{eq:Ce_R} and~\eqref{eq:psrcRXDiag} on the extinction cross sections for dielectric obstacles having relative permittivity $\Re\{\epsilonr\}\in\{10,20\}$ and $\Im\{\epsilonr\}\in\{10^{-3},10^{-5}\}$ supported in a spherical region with radius $a$.}
\label{fig:DielSphboundska}
\end{figure}

Dielectric materials (such as silicon, silica, or germanium) offer much lower losses than gold and could hence provide larger cross sections as suggested from the bounds in Fig.~\ref{fig:spherebound}. In Fig.~\ref{fig:DielSphboundska}, bounds~\eqref{eq:Ce_R} and~\eqref{eq:psrcRXDiag} on the extinction cross section are depicted for obstacles supported in a spherical region with radius $a$ and composed of a material with relative permittivity $\epsilonr=\{10+\iu 10^{-3},10+\iu 10^{-5},20+\iu 10^{-3},20+\iu 10^{-5}\}$. Here, we observe that $\CeZ$ follows~$\CeR$ for electrically larger $ka\geq 1$ cases but has a sharp drop around~$ka=1$. Effects of the losses $\Im\{\epsilonr\}$ are also relatively small for $ka\geq 1$ and negligible for $ka<1$.    

\begin{figure}[t]
\centering
\includegraphics[]{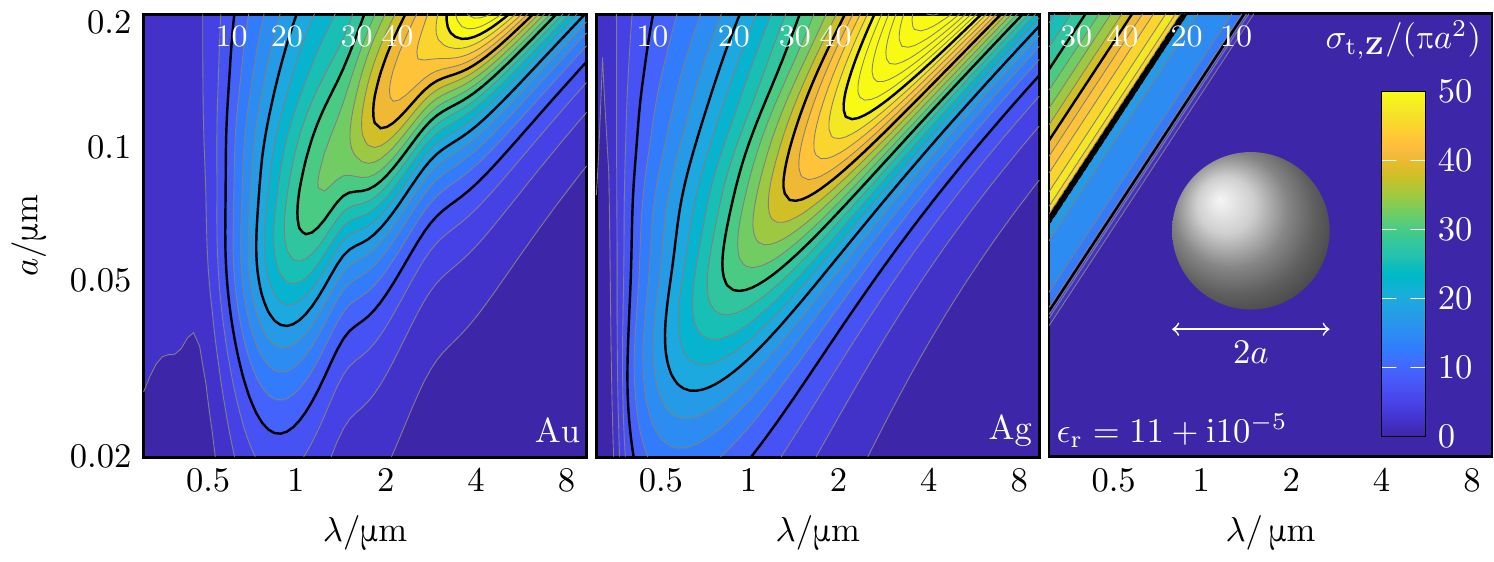}
\caption{Upper bounds on the extinction cross sections~$\CeZ/(\pi a^2)$ using~\eqref{eq:psrcRXDiag} for Au, Ag, and low-loss dielectric ($\epsilon_{\mrm{r}}=11+\iu 10^{-5}$) obstacles circumscribed by a sphere with radius~$a\in[0.02,0.2]\unit{\micro m}$. Contour lines with steps~$2.5$ are used and $\CeZ/(\pi a^2)\in\{10,20,30,40\}$ are emphasized.}
\label{fig:CeZbounds}
\end{figure}

 Contour plots depicting the upper bound~\eqref{eq:psrcRXDiag} on the normalized extinction cross section for obstacles supported within a sphere with radius $a\in[0.05,0.2]\unit{\micro m}$ and made of gold (Au), silver (Ag), and a low loss dielectric ( $\epsilon_{\mrm{r}}=11+\iu 10^{-5}$) are shown in Fig.~\ref{fig:CeZbounds}. The results show which combination of size $a$ and wavelength $\lambda$ should be used to maximize  
 $\CeZ/(\pi a^2)$. For visible light $\CeZ/(\pi a^2)$
 is limited by 20 for Au and sizes $a\in[50,100]\unit{nm}$ at the longer wavelength $750\unit{n m}$ and 5 for shorter wavelengths $400\unit{nm}$. Silver (Ag) can potentially have higher values with bounds around 30 and 10 at the longer and shorter wavelengths, respectively. The values are higher for infrared wavelengths with values above 50 for $\lambda\approx 4\unit{\micro m}$ and $a\approx 0.2\unit{\micro m}$, see also Fig.~\ref{fig:Aubounds}. Low-loss dielectric materials (here $\epsilon_{\mrm{r}}=11+\iu 10^{-5}$) have very different upper bounds compared with the two metals. It has a negligible cross section for electrically small sizes $ka<1$, \cf Fig.~\ref{fig:DielSphboundska}, as is seen for $a<50\unit{nm}$ and $\lambda>300\unit{nm}$. The upper bound on the normalized extinction cross section for visible light exceeds 40 for an obstacle with radius $200\unit{nm}$.  
 
\section{Directional scattering}

\begin{figure}
    \centering
    \includegraphics[]{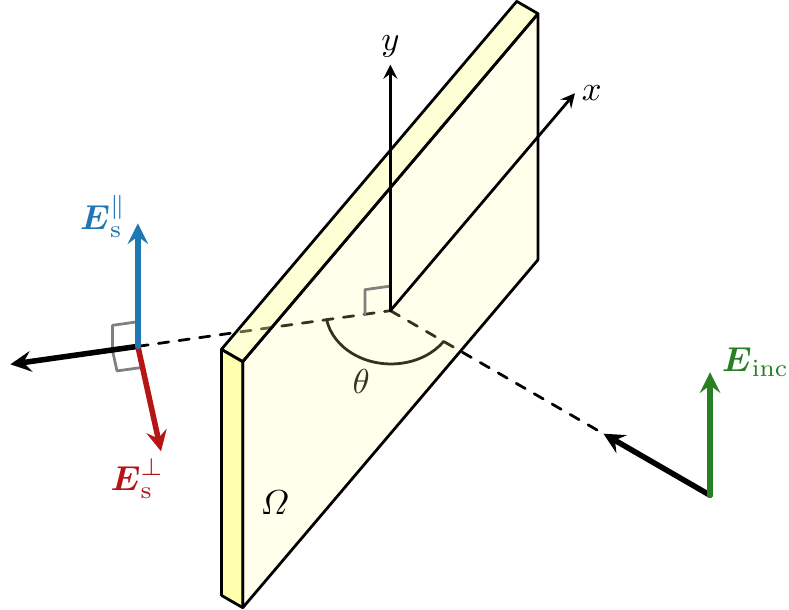}
    \caption{Schematic of a directed scattering optimization problem.  Currents within the structure $\varOmega$ are optimized, subject to material-related power balance constraints, to maximize the scattered fields in a particular polarization and direction.}
    \label{fig:directed-schem}
\end{figure}

Here we study bounds on scattering into a particular direction and polarization by a scatterer illuminated by an arbitrary incident wave. More specifically, let the excitation be described by vector~$\Vm$ and let the scattering be measured by the radiation intensity~$U( \rvh,\evh)$ in the direction~$\rvh$ with polarization~$\evh$, see Sec.~\ref{Sec:FU}. An example of this scenario where the incident field is a plane wave is depicted in Fig.~\ref{fig:directed-schem}.  In a complete analogy to Sec.~\ref{S:MaxOptR} and Sec.~\ref{S:MaxOptXR} an optimization problem is formulated for upper bounds on radiation intensity~$U(\rvh,\evh)$ using the power constraints~\eqref{Eq:Pre:Conserv2} and~\eqref{Eq:Pre:Conserv3}. The first constraint would be used when only real part of material resistivity is prescribed, while both constraints should be used to enforce both real and imaginary components of the complex resistivity.  Like absorption, scattering, and extinction studied in Sec.~\ref{S:CrossSec}, under either selection of constraints, the upper bound on radiation intensity~$U(\rvh,\evh)$ may be interpreted in terms of an upper bound on bistatic radar cross section \cite{Balanis1997}
\begin{equation}
\sigma_\mathrm{bis} = \frac{4\pi U(\rvh,\evh)}{S_0},
\end{equation} 
where the incident field is a plane wave with power flux~$S_0$.

\subsection{Prescribed losses}
\label{Sec:dir-R}

In the case of resistivity~$\rhoi$ being free, maximization of radiation intensity is determined by the solution to a QCQP 
\begin{equation}
\begin{aligned}
& \maximize && \Jm^\herm \Fm \Fm^\herm  \Jm\\
& \subto &&  \Jm^\herm \Rm \Jm - \Re\{\Jm^\herm \Vm\}  = 0   .
\end{aligned}  
\label{eq:URe_QCQP}
\end{equation}
The associated dual problem reads
\begin{equation}
U_\mat{R} = \min_{\lagMula > \lagMula_1}
\frac{\lagMula^2}{8}\Vm^\herm(-\Fm\Fm^\herm + \lagMula \Rm )^{-1}\Vm,
\label{eq:URe_dualsol}
\end{equation}
where $\lagMula_1= \Fm^\herm\Rm^{-1}\Fm$. Since the matrix~$\Fm \Fm^\herm $ is of rank 1, an analytical solution to~\eqref{eq:URe_dualsol} exists, \ie{} the Sherman–Morrison–Woodbury identity~\cite{Golub+Loan2013} may be used to evaluate the matrix inverse analytically and the simplified scalar optimization problem may be solved in closed form.  The result of this procedure is given by
\begin{equation}
    U_\mat{R} 
    =\frac{1}{8}\big(\beta+\sqrt{\alpha\gamma}\big)^2,
\end{equation}
where
\begin{equation}
\alpha = \Vm^\herm\Gm\Vm,\quad\beta = |\Fm^\herm\Gm\Vm|,\quad \gamma = \Fm^\herm\Gm\Fm,\quad
\text{and }\Gm = \Rm^{-1}.
\label{eq:bistatic-terms}
\end{equation}

\subsection{Prescribed materials}
\label{Sec:dir-RX}

If the constraint on reactive power conservation~\eqref{Eq:Pre:Conserv3} is added to the optimization problem~\eqref{eq:URe_QCQP}, the optimization problem maximizing directional scattering with prescribed material properties becomes
\begin{equation}
\begin{aligned}
& \maximize && \Jm^\herm \Fm \Fm^\herm  \Jm\\
& \subto &&  \Jm^\herm \Rm \Jm - \Re\{\Jm^\herm \Vm\}  = 0 \\
	&  &&  \Jm^\herm\Xm\Jm - \Im\{\Jm^\herm \Vm\}  = 0   
\end{aligned}  
\label{eq:UZe_Q2CQP}
\end{equation}
with the accompanying dual problem
\begin{equation}
U_\mat{Z} = \min_{(\lagMula,\lagMulb)\in \Ds}
 \frac{\lagMula^2 + \lagMulb^2}{8}\Vm^\herm \left( - \Fm \Fm^\herm + \lagMula \Rm + \lagMulb \Xm \right)^{ - 1} \Vm,
\label{eq:UZe_dualsol}
\end{equation}
where $\Ds = \{(\lagMula,\lagMulb):~ - \Fm \Fm^\herm + \lagMula \Rm + \lagMulb \Xm \succ\Om\}$.
Analogously to Sec.~\ref{Sec:ext-RX}, substitution of $\lagMulb = \lagMula\lagMulb_1$ can be used to eliminate the Lagrange multiplier~$\lagMula$.  This, together with application of the Sherman–Morrison–Woodbury identity, yields the simplified optimization problem
\begin{equation}
    U_\mat{Z} 
     =%\min_{+,-}
     \min_{\lagMulb_1\in \Ds_{1}} \frac{1+\lagMulb_1^2}{8}\big(\beta+\sqrt{\alpha\gamma}\big)^2,
    \label{eq:UZe_dualsol_simplified}
\end{equation}
where $\alpha$, $\beta$, and $\gamma$ are the same as in \eqref{eq:bistatic-terms} except for the added dependence on the Lagrange parameter $\lagMulb_1$ via the altered form of the matrix $\Gm = \left(\Rm+\lagMulb_1\Xm\right)^{-1}$. The domain~$\Ds_{1}$ is given by the union of domains~$\Ds_{1+}$ and~$\Ds_{1-}$ from~\eqref{eq:CeZrange}, \ie{}
\begin{equation}
\Ds_{1} =
    \begin{cases}        
[-( \max \lambda_n )^{-1}, -( \min \lambda_n )^{-1}] & \Xm \ \text{indefinite},\\
\R\setminus [-( \min \lambda_n )^{-1}, -( \max \lambda_n )^{-1}] & \Xm \ \text{definite}.
\end{cases}
\label{eq:UZrangeX}
\end{equation}

The inverse in~$\Gm$ may be conveniently factored using the characteristic-mode-like eigenvalue problem \eqref{eq:cm} such that each term in \eqref{eq:bistatic-terms} may be calculated efficiently using
\begin{equation}
    \mathbf{a}^\herm\Gm\mathbf{b} = \mathbf{a}^\herm\mat{Q}\left(\Id+\mu_1\boldsymbol{\Lambda}\right)^{-1}\mat{Q}^\herm\mathbf{b} = \sum_{n=1}^{N} \frac{\tilde{a}_n^*\tilde{b}_n}{1+\mu_1\lambda_n}.
\end{equation}
where $\{\lambda_n\}$ are again the eigenvalues of the characteristic-mode-like problem in \eqref{eq:cm} and the tilde denotes the projection $\mathbf{\tilde{a}} = \mat{Q}^\herm \mathbf{a}$. 

\subsection{Numerical examples}

As a simple example, consider the directive scattering bounds for the rectangular prism (slab) in Fig.~\ref{fig:directed-schem}.  The object's aspect ratio $x:y:z$ is $10:5:1$.  Illumination is a $\yvh$-polarized plane wave incident from the $\zvh$ direction.  Observation directions are restricted to the $xz$ plane and described by the angle $\theta$.  Two observation polarizations are selected: parallel ($\parallel$, electric field $\yvh$-directed) and perpendicular ($\perp$, electric field in the $xz$ plane).

\begin{figure}[tp]
    \centering
    \includegraphics[]{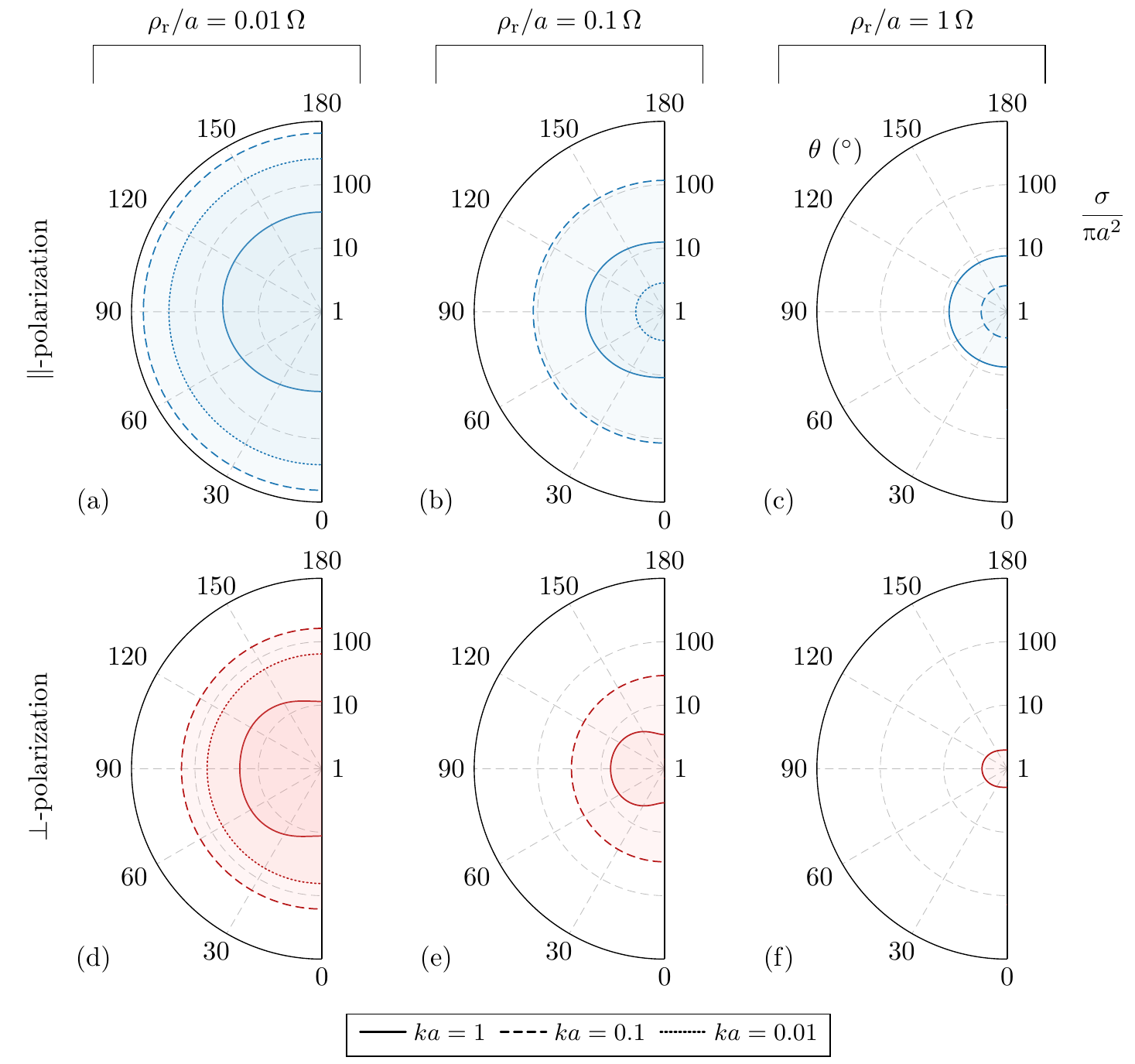}
    \caption{Upper bounds on directed scattering, plotted in terms of normalized bistatic radar cross section $\sigma_\mathrm{bis}/(\pi a^2)$, for the scattering geometry in Fig.~\ref{fig:directed-schem} with prescribed losses.}
    \label{fig:bcs_r}
\end{figure}

First, we calculate the bounds for directive scattering with prescribed losses for slabs of three electrical sizes $ka = \left\{0.01,0.1,1\right\}$ and two real resistivities $\rhor/a = \left\{0.01, 1\right\}\,\Omega$ with only the real power constraint enforced.  Results in Fig.~\ref{fig:bcs_r}, presented in terms of bistatic radar cross sections, show similar trends as observed for scattering cross sections in Fig.~\ref{fig:spherebound}, with non-monotonic shifting of the overall maximal scattering response over all observation angles as a function of electrical size.  We observe that both endfire ($\theta = \pi/2$) and broadside ($\theta = 0,2\pi$) may serve as the direction with highest scattering potential, though for small electrical sizes the angular variation in maximal scattering disappears as the optimal current becomes a uniform dipole moment.

\begin{figure}[thp]
    \centering
    \includegraphics[]{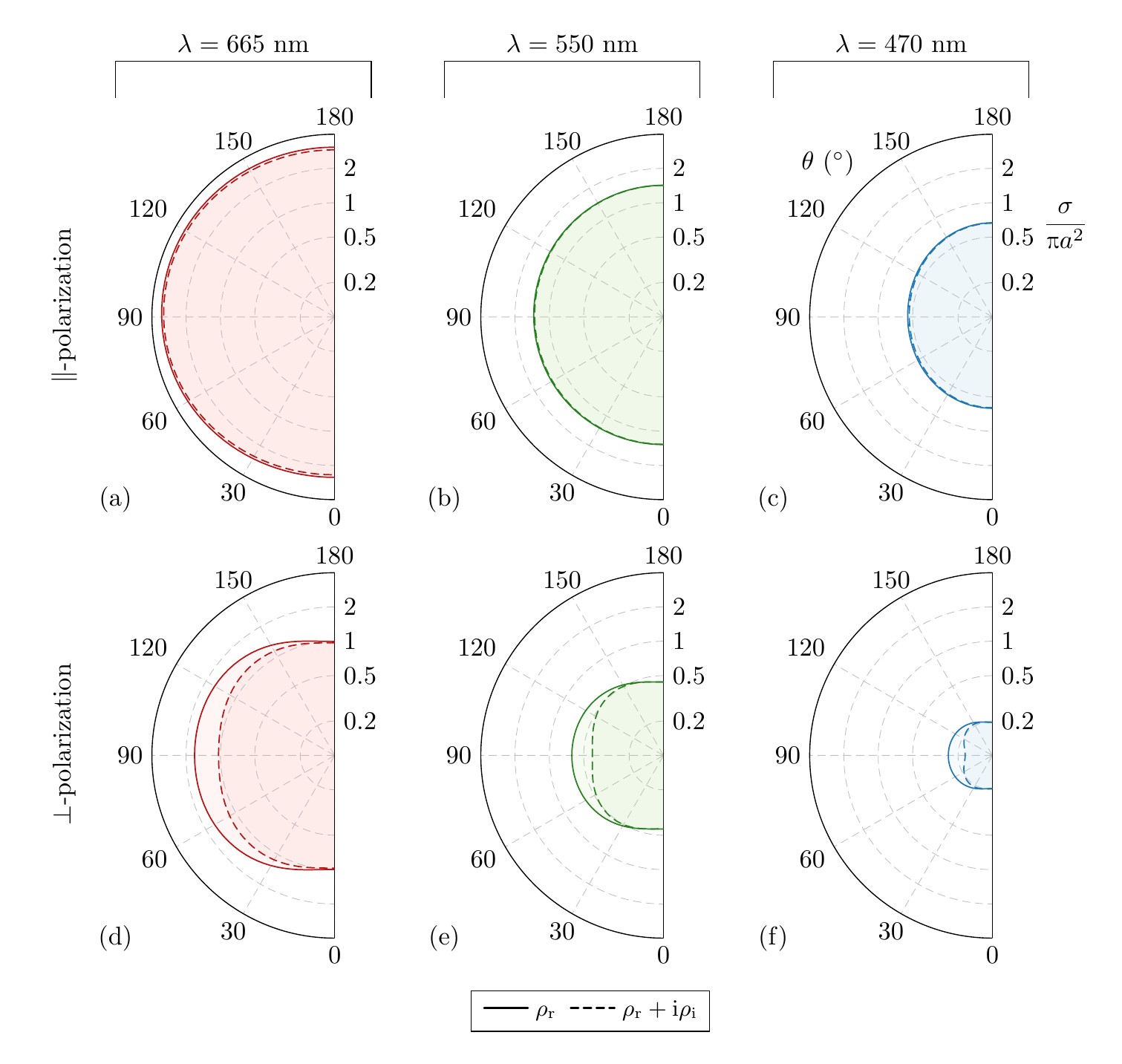}
    \caption{Directed scattering bounds for parallel (top) and perpendicular (bottom) polarizations for a gold slab at three optical wavelengths.}
    \label{fig:bcs_z}
\end{figure}

Using~\eqref{eq:UZe_dualsol_simplified}, we calculate the directive scattering bounds for the same slab with prescribed material properties corresponding to gold.  Absolute units are required in this case, so the thickness of the slab is set at~$40 \unit{nm}$ and three wavelengths $\lambda = \left\{ 470,550,665\right\}\unit{nm}$ within the optical regime are examined.  The difference in wavelength in this case is small, nonetheless the dispersive properties of gold lead to interesting variation in the maximal scattering properties, shown in Fig.~\ref{fig:bcs_z}.  We observe that over this range, the parallel polarization ($\parallel$) transitions from a maximum in the endfire direction ($\theta = 0$, $\lambda = 665\unit{nm}$) to a maximum in the broadside direction ($\theta = \pi/2$, $\lambda = 470\unit{nm}$).  Additionally, in the parallel polarization the effect of the loss versus material constraints is minimal, as also observed in the trends for extinction cross section on large spherical scatterers in this range of frequencies, \cf Fig.~\ref{fig:Aubounds}.  This is not the case, however, for the perpendicular polarization, where the inclusion of the reactance constraint impedes endfire scattering at all three wavelengths.

\section{Purcell's factor}

Upper bounds on enhancement of radiation from a point electric dipole (Purcell's factor, see Sec.~\ref{Sec:PurcellDef}) are studied in this section for the case of prescribed losses and for the case of prescribed materials.

\subsection{Prescribed losses}
\label{Sec:pur-R}
Assuming a fixed excitation vector~$\Vm$ which is the projection~\eqref{Eq:Vvec} of the electric field generated by the dipole~$\boldsymbol{p}$ in free space onto the current basis~\eqref{Eq:Current:Expan}, a maximum Purcell's factor~\eqref{Eq:Purcell:Def} with prescribed material losses is generated by 
\begin{equation}
\begin{aligned}
	& \maximize && - \Jm^\herm \Rml \Jm + \Re \{ \Jm^\herm \mat{N} \}\\
	& \subto &&  \Jm^\herm \Rm \Jm - \Re\{\Jm^\herm \Vm\}  = 0   
\end{aligned}  
\label{eq:PF_QCQP}
\end{equation}
with the corresponding dual problem
\begin{equation}
  F_\mat{R} = \min_{\lagMula > \lagMula_1}
 1 + \frac{1}{8 P_{\rad,p}}\left(\mat{N} + \lagMula\Vm \right)^\herm\big(\Rml+\lagMula \Rm \big)^{-1} \left(\mat{N} + \lagMula\Vm \right),
\label{eq:PF_dualsol}
\end{equation}
where $\lagMula_1=-1/(1+\radm_1)$ with~$\radm_1$ being the largest radiation mode eigenvalue generated by~\eqref{eq:radmodes}.

Employing diagonalization of the underlying matrices, the dual problem can be, similarly to Sec.~\ref{Sec:abs-R}, written as
\begin{equation}
F_\mat{R} = \min_{\lagMula > \lagMula_1}
 1 + \frac{1}{8 P_{\rad,p}} \sum\limits_n \frac{ \big| {\tilde N}_n + \lagMula {\tilde V}_n \big|^2 } {  1 + \lagMula \left( 1 + \varrho_n \right) }.
\label{eq:PF_dualsolDiag}
\end{equation}

\begin{figure}[t]
    \centering
    \includegraphics[]{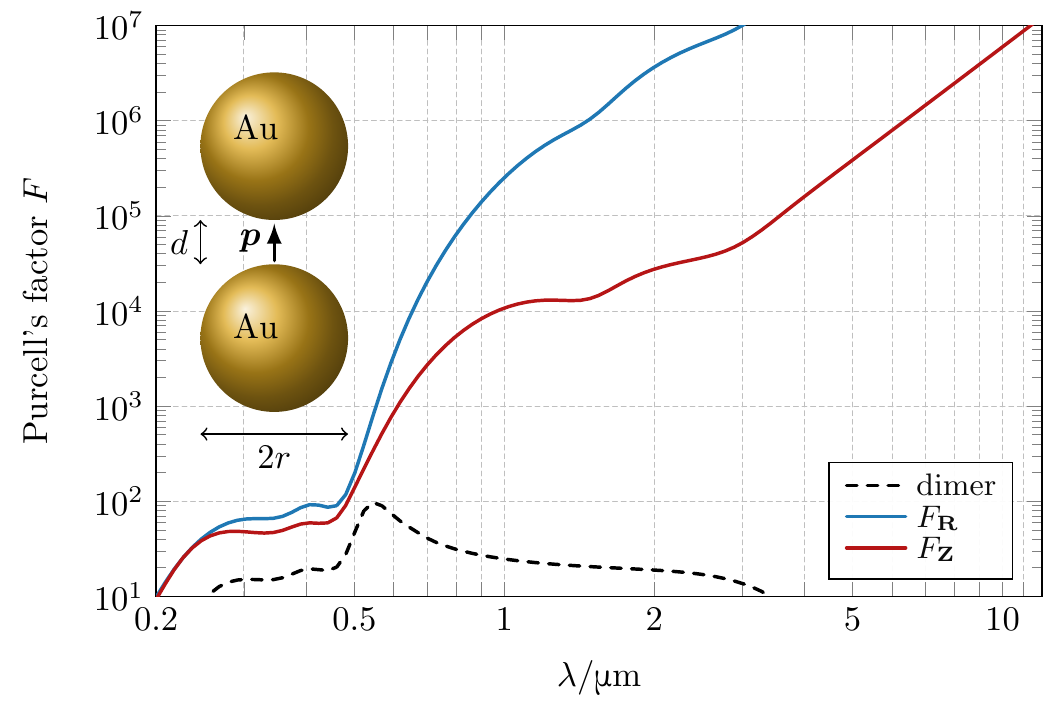}
    \caption{Frequency dependence of Purcell's factor~\eqref{Eq:Purcell:Def} and its upper bounds~\eqref{eq:PF_dualsolDiag}, \eqref{eq:PF_dualsol2} for a spherical dimer with sphere radius~$r = 30 \unit{nm}$, separation distance between the spheres~$d = 0.4 r$ and an electric dipole moment centered between the spheres and oriented along the axis of the dimer. Gold has been used for the dimer, see App.~\ref{S:MaterialModels}.}
    \label{fig:PF1}
\end{figure}

A sweep of the upper bound to Purcell's factor~$F_\mat{R}$ over frequency is shown in Fig.~\ref{fig:PF1} for a gold spherical dimer together with the realized Purcell's factor~$F$ for the same structure. A spherical geometry is used as a prototype of a core-shell particle which is commonly used for the metal-enhanced fluorescence~\cite{2010_Geddes_MetalEnhancedFluorescence, 2012_GoldNanoparticles_LouisPluchery}. In these cases, either the core or the shell are made of a plasmonic material (\eg{} gold). Within the optimization paradigm used here, the contrast current chooses the preferred option.

Comparison of both lines in~Fig.~\ref{fig:PF1} shows that in the real scenario, the dimer structure is effectively excited only in the vicinity of plasmonic resonance of the dimer and even there the realized Purcell's factor is one order in magnitude lower than upper bound~$F_\mat{R}$. Away from the plasmonic resonance a simple spherical dimer is far from acting as an optimal structure for radiation enhancement, as evidenced by a considerable gap between the realized Purcell’s factor and its upper bound.

\subsection{Prescribed material}
\label{Sec:pur-RX}
The addition of a reactive power constraint to~\eqref{eq:PF_QCQP} results in a maximal Purcell's factor under prescribed material given by the  optimization problem
\begin{equation}
\begin{aligned}
& \maximize && - \Jm^\herm \Rml \Jm + \Re \{ \Jm^\herm \mat{N} \}\\
& \subto &&  \Jm^\herm \Rm \Jm - \Re\{\Jm^\herm \Vm\}  = 0 \\
	&  &&  \Jm^\herm\Xm\Jm - \Im\{\Jm^\herm \Vm\}  = 0   
\end{aligned}  
\label{eq:PF_Q2CQP}
\end{equation}
and dual problem
\begin{equation}
F_\mat{Z} = \min_{(\lagMula,\lagMulb)\in \Ds} 1 + \frac{1}{8 P_{\rad,p}}\big(\mat{N} + ( \lagMula - \iu \lagMulb ) \Vm \big)^\herm\big(\Rml+\lagMula \Rm + \lagMulb \Xm \big)^{-1} \big(\mat{N} + ( \lagMula - \iu \lagMulb ) \Vm \big),
\label{eq:PF_dualsol2}
\end{equation}
where $\Ds = \{(\lagMula,\lagMulb):~ \Rml + \lagMula \Rm + \lagMulb \Xm \succ\Om\}$. The solution to this optimization problem must be performed by general purpose solvers, see~App.~\ref{APP:Q2CQP}.

The frequency dependence of the upper bound~$F_\mat{Z}$ is depicted in~Fig.~\ref{fig:PF1}. As compared to upper bound~$F_\mat{R}$, the additional constraint on the conservation of reactive power considerably tightens the bound at long wavelengths (small electrical sizes), pushing it closer to the realized Purcell's factor~$F$. The noticeable gap between the realized Purcell's factor and its bound nevertheless still exists away from the plasmonic resonance.  Thus, away from plasmonic resonance the realized structure would have to be modified in topology in order to achieve resonance coupling to the exciting dipole.

\section{Conclusions}

In this paper we have laid out a general optimization framework for determining bounds on several metrics related to electromagnetic scattering, absorption, and field enhancement.  In all cases, bounds were formulated using contrast current density representing the optimization domain and matrix operators mapping current distributions onto physical quantities.  Techniques such as modal decomposition and rank-1 inverse updates were applied to simplify each problem based on its distinguishing features.  Through that process, many of the problems studied here may be solved in computationally efficient ways.

The framework discussed here is not limited to the problems studied in this paper.  Further bounds on both near- and far-field metrics may be derived given the ability to calculate the necessary matrix operators.  Though in this paper all scattering objects were considered to exist in vacuum, no substantial modification is necessary for the case of objects suspended within any other lossless dielectric background.  Similarly, material inhomogeneity and anisotropy may be introduced with minimal technical changes.

The bounds presented in this paper represent the absolute optimal behavior achievable by objects of prescribed material and bounding support.  In addition to providing insight into the fundamental physical limitations of specific electromagnetic processes, these bounds will serve as benchmarks for future topology optimization implementations.  There, the associated optimal current solutions may also find utility in accelerating and informing the chosen topology optimization algorithm. 

\section*{Acknowledgement}
%We would like to thank Mats Gustafsson (Lund University), Kurt Schab (Santa Clara University), Lukas Jelinek (Czech Technical University in Prague), and Miloslav Capek (Czech Technical University in Prague) for the development of the theory of this paper, for the preparation of the text, and for generating the results. 
We would like to acknowledge the financial support of this work by the Swedish Research Council and by the Czech Science Foundation under project \mbox{No.~19-06049S}.

\appendix

\section{Power balance constraint}
\label{S:PowerBalance}

A scattering problem is typically formulated using an impressed source~$\Jv_\mathrm{i}$ which excites an incident electromagnetic field~$\vec{E}_\mathrm{i},\vec{H}_\mathrm{i}$ interacting with a scatterer, generating scattered field~$\vec{E}_\mathrm{s},\vec{H}_\mathrm{s}$. The total electromagnetic field is defined as~$\vec{E} = \vec{E}_\mathrm{i} + \vec{E}_\mathrm{s}$ and~$\vec{H} = \vec{H}_\mathrm{i} + \vec{H}_\mathrm{s}$.

A complex power balance in this scenario is typically formulated as~\cite{Jackson1999}
\begin{equation}
\label{eq:Poynting1}
- \int\limits_V \vec{E} \cdot \vec{J}_\mathrm{i}^* \mathrm{d} V  = \oint\limits_{\partial V} \left( \vec{E} \times \vec{H}^* \right) \cdot \mathrm{d} \vec{S} - \iu \omega \int\limits_V \left( \mu_0 \left| \vec{H} \right|^2 - \epsilon_0 \left| \vec{E} \right|^2 \right) \mathrm{d}V 
 + \int\limits_V \dfrac{1}{\rho^*} \left| \vec{E} \right|^2 \mathrm{d}V,
\end{equation}
see~Fig.~\ref{fig:radVSscat1} for definition of the region~$V$ and its boundary~$\partial V$.

Using a volume equivalence principle~\cite{Balanis2012}, the same scattering problem can be modeled by a contrast current density~$\Jv$ which replaces the scatterer and is the source of scattered fields. In this point of view, the complex power balance is more naturally stated as
\begin{equation}
\label{eq:Poynting2}
- \int\limits_V \vec{E} \cdot \left( \Jv + \vec{J}_\mathrm{i} \right)^* \mathrm{d} V  = \oint\limits_{\partial V} \left( \vec{E} \times \vec{H}^* \right) \cdot \mathrm{d} \vec{S} - \iu \omega \int\limits_V \left( \mu_0 \left| \vec{H} \right|^2 - \epsilon_0 \left| \vec{E} \right|^2 \right) \mathrm{d}V,
\end{equation}
where sources~$\Jv + \vec{J}_\mathrm{i}$ radiate in free space.

Comparing~\eqref{eq:Poynting1} and~\eqref{eq:Poynting2} identifies the term 
\begin{equation}
\label{eq:Pzmat}
\int\limits_V \vec{E} \cdot \Jv^* \mathrm{d} V  = \int\limits_V \dfrac{1}{\rho^*} \left| \vec{E} \right|^2 \mathrm{d}V = \int\limits_V \rho \left| \vec{J} \right|^2 \mathrm{d}V \approx \Jm^\herm \left(\Rml + \iu\Xmm \right)\Jm
\end{equation}
with a complex power flow within the material of the scatterer. The last equality results from a current expansion~\eqref{Eq:Current:Expan} and definitions introduced in Sec.~\ref{Sec:ContrastCurr} and defines the matrix
\begin{equation}
\label{eq:PzmatExpl}
\left(\Rml + \iu\Xmm \right)_{ij}  =  \int\limits_V  \psiv_i^* \cdot \rho \, \psiv_j \diff V.
\end{equation}

Within the volume equivalence, the scattered field is itself a valid solution to Maxwell's equations in free space and therefore generates a complex power balance relation
\begin{equation}
\label{eq:Poynting3}
- \int\limits_V \vec{E}_\mathrm{s} \cdot \vec{J}^* \diff V  = \oint\limits_{\partial V} \left( \vec{E}_\mathrm{s} \times \vec{H}_\mathrm{s}^* \right) \cdot \diff \vec{S} - \iu \omega \int\limits_V \left( \mu_0 \left| \vec{H}_\mathrm{s} \right|^2 - \epsilon_0 \left| \vec{E}_\mathrm{s} \right|^2 \right) \diff V,
\end{equation} 
where
\begin{equation}
\label{eq:Pzvac}
- \int\limits_V \vec{E}_\mathrm{s} \cdot \vec{J}^* \diff V \approx \Jm^\herm \left(\Rmr + \iu\Xm_0 \right)\Jm
\end{equation}
defines matrix~$\Rmr - \iu\Xm_0$ introduced in Sec.~\ref{Sec:PowerConst} as a projection of scattered field operator
\begin{equation}
    \label{eq:EsJ}
    \vec{E}_\mathrm{s} \left( \Jv  \right) = \iu k \eta_0\int_V\left(\mat{1}+k^{-2}\nabla\nabla\right)\cdot\Jv\left(\rv'\right)\frac{\displaystyle \eu^{\iu k\lvert\rv-\rv'\rvert}}{\displaystyle 4\pi\lvert\rv-\rv'\rvert} \diff V,
\end{equation}
onto a current basis~\eqref{Eq:Current:Expan}
\begin{equation}
\label{eq:PzvacExpl}
(\Rmr + \iu\Xm_0)_{ij}  =  \int_V \psiv_i^* \cdot \vec{E}_\mathrm{s} (\psiv_j) \diff V,
\end{equation}
see~\cite{Harrington1968} for more details.

Addition of~\eqref{eq:Pzmat} and~\eqref{eq:Pzvac} together with definition~\eqref{Eq:Vvec} generates the complex power constraint~\eqref{eq:poynting}.

\section{Quadratically constrained quadratic programs}
\label{APP:Q2CQP}
Assume an optimization problem of the form
\begin{equation}
\begin{aligned}
	& \maximize && \Jm^\herm \mat{A} \Jm + \Re \left\{ \Jm^\herm \mat{a} \right\} + a_0\\
	& \subto &&  \Jm^\herm \mat{B} \Jm + \Re \left\{ \Jm^\herm \mat{b} \right\} + b_0 = 0\\
	&  &&  \Jm^\herm \mat{C} \Jm + \Re \left\{ \Jm^\herm \mat{c} \right\} + c_0 = 0
\end{aligned}  
\label{eq:Q2CQP}
\end{equation}
where $\mat{A} = \mat{A}^\herm,\mat{B} = \mat{B}^\herm, \mat{C} = \mat{C}^\herm \in \mathbb{C}^{\left[N \times N\right]}$, $a_0,
b_0, c_0 \in \mathbb{R}$ and $\mat{a},\mat{b}, \mat{c},\Jm \in \mathbb{C}^{\left[N \times 1\right]}$.  The dual function~\cite{Boyd+Vandenberghe2004,Park+Boyd2017} for this problem is
\begin{equation}
g \left( \lagMula, \lagMulb \right)
=-\frac{1}{4}(\mat{a}-\lagMula \mat{b}-\lagMulb \mat{c})^{\herm} \mat{H}^{-1}(\mat{a}-\lagMula\mat{b}-\lagMulb\mat{c}) + a_0-\lagMula b_0-\lagMulb c_0,
\label{eq:Q2CQPDual}
\end{equation}
where the Hessian matrix of the Lagrangian (strictly speaking, twice the Hessian matrix)
\begin{equation}
\mat{H} = \mat{A}-\lagMula\mat{B}-\lagMulb\mat{C}
\label{eq:Q2CQPHess}
\end{equation}
is assumed to be negative definite, \ie{} $\mat{H} \prec\Om$, and where $\lagMula,\lagMulb \in \mathbb{R}$ are Lagrange multipliers.  The stationary point of the Lagrangian is found at
\begin{equation}
\mat{I}_\mrm{opt}
= - \frac{1}{2} \mat{H}^{-1}(\mat{a}-\lagMula\mat{b}-\lagMulb\mat{c}).
\label{eq:Q2CQPIstat}
\end{equation}

The dual solution to the problem~\eqref{eq:Q2CQP} is realized by minimizing the convex dual function~$g \left( \lagMula, \lagMulb \right)$. Since the primal optimization problem~\eqref{eq:Q2CQP} is generally not convex~\cite{Boyd+Vandenberghe2004}, the solution to a dual problem in general only gives an upper bound to the primal problem. However, the case with a single quadratic constraint ($\mat{C}=\mat{0}$) in~\eqref{eq:Q2CQP} is solved by the dual formulation if the solution is feasible~\cite{Boyd+Vandenberghe2004}. Strong duality also holds for two quadratic constrains under mild conditions on $\mat{B}$ and $\mat{C}$~\cite{Beck+Eldar2006}.
In all cases treated in this paper, there seems to be (inductive observation based on many numerical trials) no dual gap~\cite{Boyd+Vandenberghe2004} and the solution to the dual problem equals the solution to the primal problem.

The minimization of~\eqref{eq:Q2CQPDual} must typically be done via general purpose convex optimization tools. Due to its simplicity and fast convergence, Newton's method is often the first choice~\cite{Nocedal+Wright2006}. Another justification for the use of Newton's method is the existence of simple formulas for the first and the second derivatives of the dual function~$g \left( \lagMula, \lagMulb \right)$ with respect to the Lagrange multipliers.  These derivatives are given by
\begin{align}
\label{Eq:Q2CQPdgdL1}
\frac{\partial g \left( \lagMula, \lagMulb \right)}{\partial \lagMula} &= - \Re \left\{ \left( \mat{B} \mat{I}_\mrm{opt} + \mat{b} \right)^\herm \mat{I}_\mrm{opt} \right\} - {b_0} \\
\frac{\partial g \left( \lagMula, \lagMulb \right)}{\partial \lagMulb} &= - \Re \left\{ \left( \mat{C} \mat{I}_\mrm{opt} + \mat{c} \right)^\herm \mat{I}_\mrm{opt} \right\} - {c_0} \\
\frac{\partial^2 g \left( \lagMula, \lagMulb \right)} {\partial \lagMula^2} &=  - 2 \left( \mat{B} \mat{I}_\mrm{opt} + \frac{\mat{b}}{2} \right)^\herm \mat{H}^{-1} \left( \mat{B} \mat{I}_\mrm{opt} + \frac{\mat{b}}{2} \right) \\
\frac{\partial^2 g \left( \lagMula, \lagMulb \right)} {\partial \lagMulb^2} &=  - 2 \left( \mat{C} \mat{I}_\mrm{opt} + \frac{\mat{c}}{2} \right)^\herm \mat{H}^{-1} \left( \mat{C} \mat{I}_\mrm{opt} + \frac{\mat{c}}{2} \right) \\
\frac{\partial^2 g \left( \lagMula, \lagMulb \right)} {\partial \lagMula \partial \lagMulb} &=  - 2 \Re \left\{ \left( \mat{B} \mat{I}_\mrm{opt} + \frac{\mat{b}}{2} \right)^\herm \mat{H}^{-1} \left( \mat{C} \mat{I}_\mrm{opt} + \frac{\mat{c}}{2} \right) \right\}
\end{align}

\subsection{Single quadratic constraint}
\label{App:Q1CQP}
If~$\mat{C},\mat{c},c_0 = 0$ in the optimization problem~\eqref{eq:Q2CQP}, considerable simplifications may be made. For example, assume that a matrix~$\mat{Q}$ exists that simultaneously diagonalizes the matrices~$\mat{A},\mat{B}$. The optimization problem~\eqref{eq:Q2CQP} can then be recast into an alternative (``$\mat{Q}$'') basis as
\begin{equation}
\begin{aligned}
& \maximize && {\tilde \Jm}^\herm \mat{\tilde A} {\tilde \Jm} + \Re \big\{ {\tilde \Jm}^\herm \mat{\tilde a} \big\} + a_0\\
& \subto &&  {\tilde \Jm}^\herm \mat{\tilde B} {\tilde \Jm} + \Re \big\{ {\tilde \Jm}^\herm \mat{\tilde b} \big\} + b_0 = 0  
\end{aligned}  
\label{eq:Q1CQPTilde}
\end{equation}
where $\mat{\Jm} = \mat{Q \tilde \Jm}, \mat{\tilde a} = \mat{Q}^\herm \mat{a}, \mat{\tilde b} = \mat{Q}^\herm \mat{b}$ and where the matrices~$\mat{\tilde A} = \mat{Q}^\herm \mat{AQ}, \mat{\tilde B} = \mat{Q}^\herm \mat{BQ}$ are real and diagonal. This leads to an analytical inversion of the Hessian matrix
\begin{equation}
\label{Eq:Q1CQPinvH}
{\tilde H}_{nn}^{-1} = \big( {\tilde A}_{nn} - \lagMula {\tilde B}_{nn} \big)^{-1},
\end{equation}
which in many cases allows for an analytical solution to the problem~\eqref{eq:Q1CQPTilde} that is given by the root of~\eqref{Eq:Q2CQPdgdL1}.

For two important cases considered in this paper, an analytical solution to~\eqref{eq:Q1CQPTilde} can be found. The first case occurs when $\mat{A} = 0$ and the matrix~$\mat{B}$ is of full rank and not indefinite. The $\mat{Q}$-basis is in that case formed by an eigenvalue decomposition $\mat{BQ} = \mat{QD}$ and the Lagrange multiplier which minimizes the dual function is given by
\begin{equation}
\lagMula_\mathrm{opt}^2 = \frac{ \sum_n  \left| {\tilde a}_n \right|^2 {\tilde B}_{nn}^{-1}  } { \sum\limits_n  \left( \big| {\tilde b}_n \big|^2 {\tilde B}_{nn}^{-1} \right)   - 4b_0}
\label{Eq:Q1CQPA0lambda}
\end{equation}
under the condition that~$\lagMula_\mathrm{opt}^2 > 0$. The positive root of~\eqref{Eq:Q1CQPA0lambda} is chosen when~${\tilde B}_{nn} > 0$, while the negative root of~\eqref{Eq:Q1CQPA0lambda} is chosen when~${\tilde B}_{nn} < 0$.

The second case is realized when the matrix~$\mat{A} = \alpha \mat{L}_1^\herm \mat{L}_1$ is a rank-1 Hermitian matrix, the matrix~$\mat{B} = \beta \mat{L}_2^\herm \mat{L}_2$ is Hermitian and of full rank, the vector~$\mat{a} = 0$ and~$b_0 = 0$. The $\mat{Q}$-basis is most easily obtained via~$\mat{Q} = \mat{L}_2^{-1} \Um_1$, where~$\Um_1$ is a unitary matrix coming from the singular value decomposition~$ \mat{L}_2^{-\herm} \mat{L}_1^\herm =  \Um_1\Sigmam\Um_2^\herm $. Since the matrix~$\mat{A}$ is rank-1, only one element of the diagonal matrix~$\mat{\tilde A}$ is non-zero. Denoting its index as~$i = 1$, the solution to this special case reads
\begin{equation}
\lagMula_\mathrm{opt} = \frac{ {\tilde A}_{11} } { {\tilde B}_{11} } \left( 1 \pm { \left( \frac{ {\tilde B}_{11} }{ \big| {\tilde b}_1 \big|^2} \sum\limits_n \frac{ \big| {\tilde b}_n \big|^2 }{ {\tilde B}_{nn}} \right)}^{ - \frac{1}{2}} \right),
\label{Eq:Q1CQPArank1a0b00}
\end{equation}
where the correct sign in~\eqref{Eq:Q1CQPArank1a0b00} is chosen according to the negative definiteness of the Hessian matrix~\eqref{eq:Q2CQPHess}.

%\bibliography{total}

\section{Radiation modes}\label{S:radmodes}
\begin{figure}
\centering
\includegraphics[width=0.7\columnwidth]{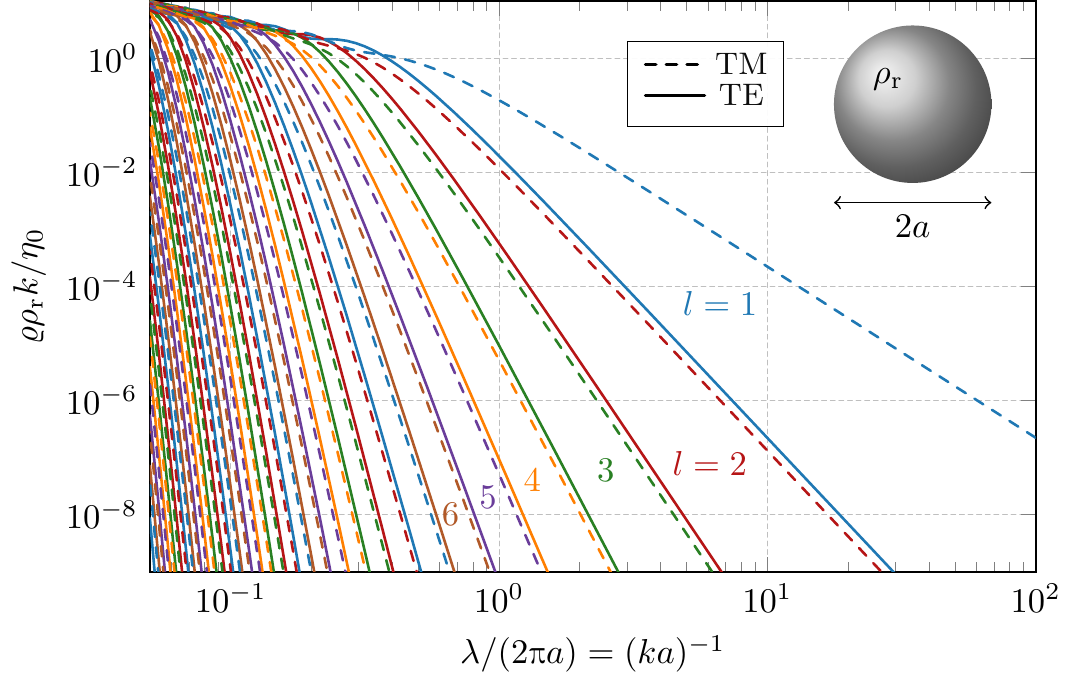}%
\caption{Normalized radiation modes, $\radm_n$, for a homogeneous spherical region with radius $a$, resistivity $\rhor$, and free-space wavenumber~$k$. Modes are ordered as $n=2(l^2+l-1+(-1)^s m+\tau$, with orders $l=\left\{1,2,\ldots\right\}$ and TE ($\tau=1$) and TM ($\tau=2$) in solid and dashed curves, respectively.}
\label{fig:Sphradmodes}
\end{figure}

Radiation modes are determined from the eigenvalue problem~\eqref{eq:radmodes}. Although radiation modes are easily determined numerically for arbitrary shapes using the $\Sm$ matrix~\cite{Tayli+etal2018}, analytic expressions are valuable. 
The radiation modes for a solid sphere with homogeneous resistivity $\rhor$ can be determined analytically
\begin{multline}
  \varrho_{n}
  = \frac{k^2\eta_0}{\rhor}\int_0^a |\uop^{(1)}_{\sphi}(kr)|^2\diffV
  =\frac{k^2\eta_0 a^3}{2\rhor}\left((\Rop^{(1)}_{1,l})^2 - \Rop^{(1)}_{1,l-1}\Rop^{(1)}_{1,l+1}
  +\frac{2}{ka}\Rop^{(1)}_{1,l}\Rop^{(1)}_{2,l}\delta_{\tau,2}\right)\\
  \approx
  \frac{(ka)^{2l}}{\big((2l+1)!!\big)^{2}}\frac{\eta_0 a}{\rhor}
  \begin{cases}
    (ka)^2/(2l) & \tau=1 \\
    (l+1) & \tau=2
  \end{cases}
  \qtext{as }
  ka\to 0,
\label{eq:radmodes_sph}
\end{multline}
where $\uop^{(1)}_{\sphi}$ denote regular spherical waves~\cite{Hansen1988,Kristensson2016}, $\Rop^{(1)}_{\tau,l}=\Rop^{(1)}_{\tau,l}(ka)$ radial functions~\cite{Hansen1988,Tayli+etal2018}, and $\sphi=(\tau,s,m,l)$ a multi-index with $s,l\in\{1,2\}$, $m\in\{0,\dots,l\}$, and $l\in\{1,\dots\}$ ordered as $n=2(l^2+l-1+(-1)^s m+\tau$. Amplitudes of radiation modes for a homogeneous sphere with radius $a$ are depicted in Fig.~\ref{fig:Sphradmodes}. The radiation modes are dominated by the three transverse magnetic (TM) ($\tau=2$) dipole modes ($l=1$) for electrically small spheres followed by the three transverse electric (TE) ($\tau=1$) dipole and five TM quadrupole ($l=2$) modes. They increase monotonically with the electrical size $ka$. The sum of radiation modes for a homogeneous object with volume $\vol$ and resistivity $\rhor$ are determined from the trace of matrix~$\Rmr$ to $\sum\radm_n=k^2\eta_0\vol/(2\pi\rhor)$.

Radiation modes can also be determined analytically in the limit of electrically small objects. In this limit, electric dipoles ($\tau=2$ and $l=1$) dominate radiation and the Rayleigh quotient related to the eigenvalue problem~\eqref{eq:radmodes} simplifies to
% \begin{equation}
%   \frac{\Jm^\herm\Rmr\Jm}{\Jm^\herm\Rml\Jm}
%   =  \frac{|\Sm\Jm|^2}{\Jm^\herm\Rml\Jm}
%   \approx
%   \frac{k^2\eta_0\sum_{\sphi}\big|\int_\reg \uop_{\sphi}^{(1)}\cdot\Jv\diffV\big|^2}{\int_\reg \rhor|\Jv|^2\diffV}
%   \approx
%   \frac{k^2\eta_0\sum_{\sphi}\left|\int_\reg \evh\cdot\Jv\diffV\right|^2}{6\pi\int_\reg \rhor|\Jv|^2\diffV},
% \label{eq:radModes1}
% \end{equation} 
\begin{equation}
\begin{split}
  \frac{\Jm^\herm\Rmr\Jm}{\Jm^\herm\Rml\Jm}
  =  \frac{|\Sm\Jm|^2}{\Jm^\herm\Rml\Jm} \approx
  \dfrac{k^2\eta_0\sum\limits_{\sphi}\Big|\displaystyle\int \uop_{\sphi}^{(1)}\cdot\,\Jv\diffV\Big|^2}{\displaystyle\int \rhor|\Jv|^2\diffV} \approx
  \dfrac{k^2\eta_0\sum\limits_{\sphi}\Big|\displaystyle\int \evh\cdot\Jv\diffV\Big|^2}{6\pi\displaystyle\int \rhor|\Jv|^2\diffV},
\label{eq:radModes1}
\end{split}
\end{equation} 
where we used the low-frequency expansion of the regular spherical waves $\uop_{\sphi}^{(1)}\approx \evh/\sqrt{6\pi}$ for electric dipoles in the $\evh\in\{\xvh,\yvh,\zvh\}$-directions.

For inhomogeneous objects with resistivity $\rhor(\rv)$, we use variational calculus and set $\Jv=\Jv_0+\delta \Jv_1$ and evaluate for small perturbations $\delta\Jv_1$ with $\delta\to 0$ giving
% \begin{equation}
% 	\frac{\left|\int_\reg \evh\cdot\Jv_0+\delta \evh\cdot\Jv_1\diffV\right|^2}{\int_\reg \rhor(|\Jv_0|^2+2\delta \Jv_1\cdot \Jv_0+\delta^2 |\Jv_1|^2)\diffV}
% 	\approx\frac{\left|\int_\reg \evh\cdot\Jv_0\diffV\right|^2+2\delta\int_{\reg} \evh\cdot\Jv_1\diffV\int_\reg \evh\cdot\Jv_0\diffV}{\int_\reg \rhor |\Jv_0|^2\diffV+2\delta\int_{\reg} \rhor \Jv_1\cdot \Jv_0\diffV}.
% \label{eq:radModes2}
% \end{equation}
\begin{equation}
\begin{split}
\dfrac{\displaystyle\Big|\int ( \evh\cdot\Jv_0+\delta \evh\cdot\Jv_1 ) \diff V\Big|^2}{\displaystyle \int \rhor(|\Jv_0|^2+2\delta \Jv_1\cdot \Jv_0+\delta^2 |\Jv_1|^2)\diff V}
\approx\dfrac{\displaystyle \Big|\int \evh\cdot\Jv_0\diff V\Big|^2+2\delta\int \evh\cdot\Jv_1\diff V\int \evh\cdot\Jv_0\diff V}{\displaystyle \int \rhor |\Jv_0|^2\diff V+2\delta\int \rhor \Jv_1\cdot \Jv_0\diff V}.
\label{eq:radModes2}
\end{split}
\end{equation}
The functional is stationary for all perturbations $\Jv_1$ if
\begin{equation}
	\int_{\reg} \evh\cdot\Jv_1\diff V = 0
	\qtext{and }
	\int_{\reg} \rhor \Jv_1\cdot \Jv_0\diff V=0 
\label{eq:radModes3}
\end{equation}
for all $\Jv_1$ which imply a current density $\Jv_0\sim \evh/\rhor$. 
Within this approximation, the first three dominant radiation modes have eigenvalues
\begin{equation}
	\radm_n = \frac{k^2\eta_0}{6\pi}\int_{\reg} \dfrac{1}{\rhor(\rv)}\diff V
  =\frac{k^2\eta_0\vol}{6\pi\rhor}
  \qtext{for } n=1,2,3
\label{eq:radmodesmall}
\end{equation}
for $ka\ll 1$ and the last equality is for homogeneous objects with volume $\vol$.

\begin{figure}[tp]
\centering
\includegraphics[]{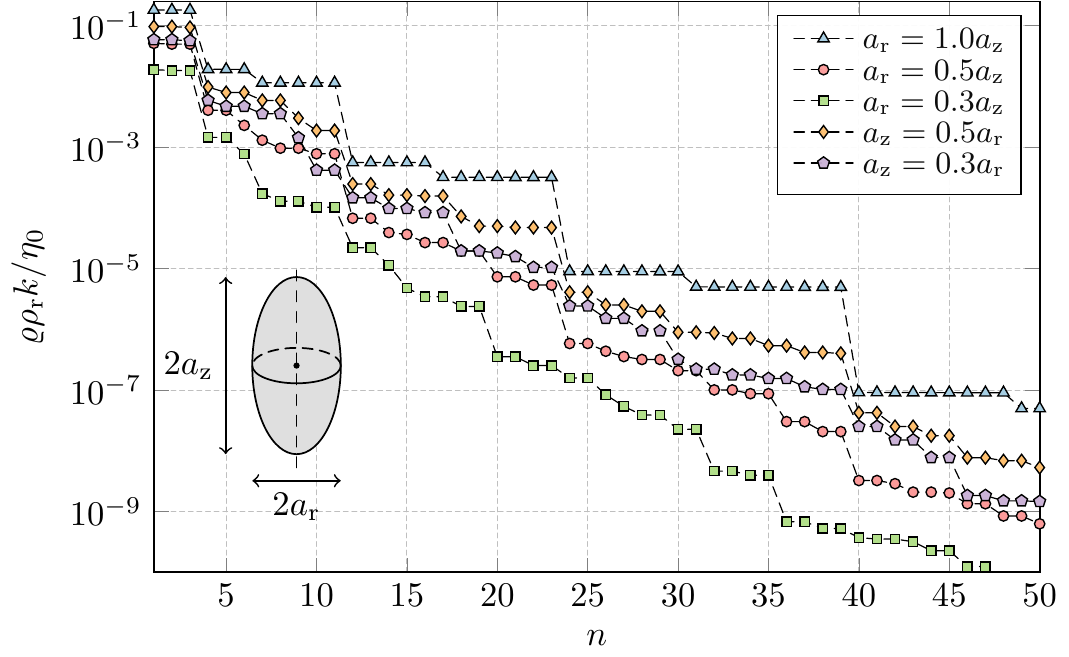}
\caption{Radiation modes $\radm_n$ for spheroids with semi-axes $a_{\mrm{z}}$ and $a_{\mrm{r}}$ for $ka=1$ with $a=\max\{a_{\mrm{z}},a_{\mrm{r}}\}$ and resistivity $\rhor$ normalized with $\eta_0/(\rhor k)$.}%
\label{fig:Sphoidradmodes}
\end{figure}

Radiation modes for a spherical and oblate and prolate spheroidal regions with semi-axes~$a_{\mrm{r}}$ and $a_{\mrm{z}}$ are depicted in Fig.~\ref{fig:Sphoidradmodes}. The first three modes radiates as electric dipoles and have eigenvalues approximately given by the volume $\vol=4\pi a^2_{\mrm{r}}a_{\mrm{z}}/3$. This explains the decrease of the first modes compared to the sphere and ordering of the curves. Eigenvalues for the sphere appear in groups with $2l+1$ elements, where $l$ denote the order of the spherical mode, see also~\eqref{eq:radmodes_sph} and Fig.~\ref{fig:Sphradmodes}. The eigenvalues decrease rapidly with the mode index such that higher order become negligible for $l\gg ka$ with $n=2l(l+2)$ and the order, $l$, can often be estimated from $l=\lceil ka+7\sqrt[3]{ka}+3\rceil$~\cite{Song+Chew2001}.    

Radiation modes together with an expansion of the incident plane wave in spherical waves~\cite{Kristensson2016} 
\begin{equation}
    a_{\sphi} = 4\pi\iu^{l-\tau+1}\evh\cdot\mat{Y}_{\sphi}(\kvh),
\end{equation}
are used in~\eqref{eq:Ca_R},~\eqref{eq:Cs_R}, and~\eqref{eq:Ce_R} to express the bounds on the cross sections in a form that is suitable for small size approximations. Here,  $\mat{Y}_{\sphi}$ denotes spherical harmonics~\cite{Kristensson2016} and the explicit value $|\evh\cdot\mat{Y}_{\sphi}|^2=3/(8\pi)$ for the dipole terms ($l=1$) \begin{equation}
  |a_1|^2=16\pi^2\frac{3}{8\pi}
  =6\pi
\label{eq:radModes4}
\end{equation}
is used in~\eqref{eq:Ca_Rlow},~\eqref{eq:Cs_Rlow}, and~\eqref{eq:Ce_Rlow}.

\section{Material models}\label{S:MaterialModels}

\begin{figure}[h!]
    \centering
    \includegraphics[]{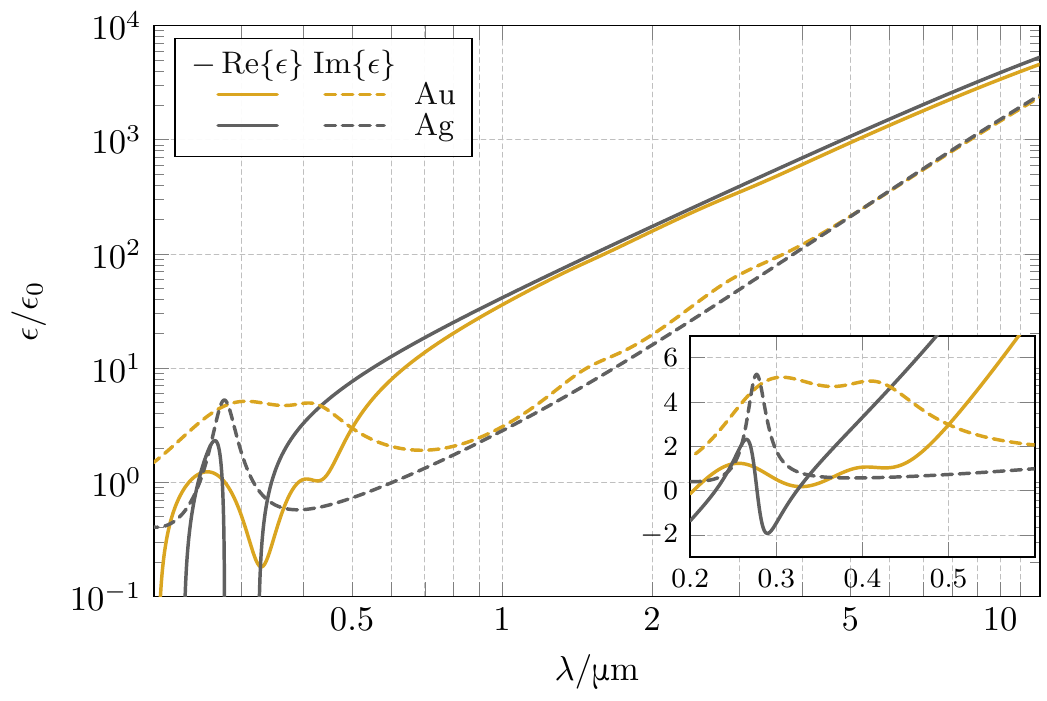}
    \caption{Frequency dependent permittivity of gold and silver. The inset shows a linear scale plot of the region with negative material parameters.}
    \label{fig:epsr}
\end{figure}

\begin{figure}[h!]
    \centering
    \includegraphics[]{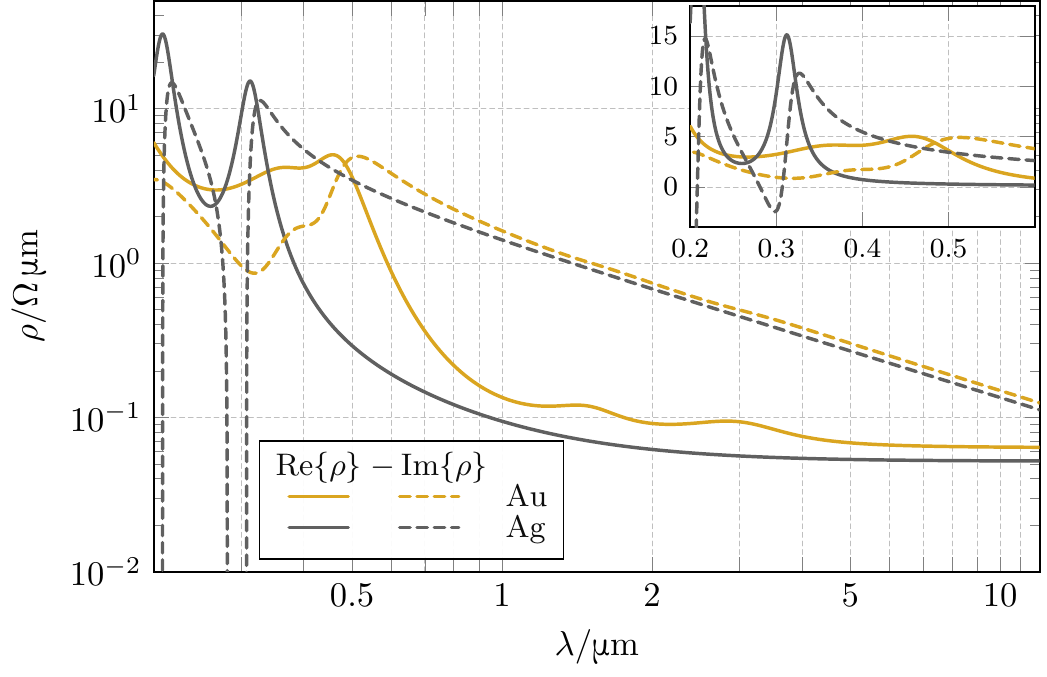}
    \caption{Frequency dependent resistivity of gold and silver. The inset shows a linear scale plot of the region with negative material parameters.}
    \label{fig:rho}
\end{figure}

A MATLAB implementation~\cite{Drude_Lorentz_at_file_exchange} of a Drude-Lorentz model of gold and silver~\cite{Rakic+etal1998} is used for the permittivity $\epsilon$ and resistivity $\rho$ throughout the paper. The frequency dependence of both parameters in the interval~$\lambda \in [0.2,12]\unit{\micro m}$ is depicted in~Fig.~\ref{fig:epsr} and Fig.~\ref{fig:rho}. This permittivity model also agrees well with older reference data for gold from~\cite{Johnson+Christy1972}.  Bulk material parameters are assumed throughout the paper. Surface effects resulting in a non-local material response are neglected.

\bibliographystyle{ieeetr}
\bibliography{total,bibadd}

\end{document}